\documentclass[journal]{IEEEtran}

\usepackage{amsmath,amssymb,amsfonts}
\usepackage{algorithm}
\usepackage{algorithmic}
\usepackage{graphicx}
\usepackage{cite}
\usepackage{booktabs}
\usepackage{multirow}
\usepackage{balance}
\usepackage{url}
\usepackage{comment}
\usepackage{enumitem}
\usepackage[table]{xcolor}  
\usepackage{multirow}  
\usepackage{threeparttable}
\usepackage[caption=false,font=footnotesize]{subfig}

\usepackage{xspace}
\newcommand{\sysname}{$\text{CORE-LEO}$\xspace}

\newif\ifshowedits
\showeditsfalse   
\newcommand{\add}[1]{\ifshowedits{\color{blue}#1}\else#1\fi}

\usepackage{xcolor}

\newif\ifshowedits
\showeditsfalse   

\usepackage[normalem]{ulem}

\def\BibTeX{{\rm B\kern-.05em{\sc i\kern-.025em b}\kern-.08em
    T\kern-.1667em\lower.7ex\hbox{E}\kern-.125emX}}

\begin{document}

\title{A RAG-Enhanced Bi-Level Cognitive Orchestration Framework for LEO Satellite Networks}

\author{Yuhong Jiang, Zhishu Shen,~\IEEEmembership{Member, IEEE}, Tong Yin, Qiushi Zheng,~\IEEEmembership{Member, IEEE}, Yichao Jin,~\IEEEmembership{Senior Member, IEEE}, Fidan Mehmeti,~\IEEEmembership{Senior Member, IEEE}, and Jiong~Jin,~\IEEEmembership{Member,~IEEE} 

\thanks{Yuhong Jiang, Zhishu Shen, and Tong Yin are with the School of Computer Science and Artificial Intelligence, Wuhan University of Technology, Wuhan, China (e-mail: kyou@whut.edu.cn, z\_shen@ieee.org, yt\_354075@whut.edu.cn). Zhishu Shen is also with the Hubei Key Laboratory of Transportation Internet of Things, Wuhan University of Technology, Wuhan, China.}
\thanks{Qiushi Zheng, and Jiong Jin are with the School of Engineering, Swinburne University of Technology, Melbourne, Australia (e-mail: \{qiushizheng, jiongjin\}@swin.edu.au)}
\thanks{Yichao Jin is with School of Automation (School of Artificial Intelligence), Hangzhou Dianzi University, Hangzhou, China (email: yichao.jin@ieee.org).} 
\thanks{Fidan Mehmeti is with the Chair of Communication Networks, Technical University of Munich, Munich, Germany (e-mail: fidan.mehmeti@tum.de).}
\thanks{This work was supported in part by the National Natural Science Foundation of China (Grant No. 62472332).}
\thanks{\textit{Corresponding author: Zhishu Shen.}}
}
\maketitle

\begin{abstract}
The rapid growth of remote sensing data in Low Earth Orbit (LEO) satellite networks is increasingly constrained by limited downlink capacity to terrestrial networks. Satellite edge computing alleviates this pressure by enabling in-orbit data processing. However, it introduces a new challenge of spatio-temporal resource fragmentation. Variations in onboard computing capability, constrained energy availability, and intermittent inter-satellite and satellite–ground connectivity lead to highly dynamic and uneven resource distribution, which degrades the performance of conventional static routing and scheduling approaches. To address this, we propose a Retrieval-Augmented Generation (RAG)-enhanced 
bi-level cognitive orchestration framework for knowledge-guided, 
multi-objective scheduling. The proposed framework explicitly decouples 
network control across two different operational scales: at the strategic 
upper level, a Large Language Model (LLM) leverages an offline-distilled 
Expert Knowledge Base (EKB) to dynamically infer preference weights based 
on a compact abstract-state descriptor of real-time network conditions. 
At the lower execution level, a fidelity-aware genetic scheduler utilizes 
these inferred weights to compute physically feasible, collision-free 
joint routing and task offloading schedules. Extensive evaluations on a 
high-fidelity Walker-Delta network testbed under mixed-criticality 
workloads demonstrate that the proposed framework effectively consolidates 
fragmented resources, achieving a 30.7\% reduction in packet loss, a 30\% 
improvement in energy efficiency over the most competitive learning-based 
baseline, and an 8.5\% decrease in end-to-end latency, while maintaining 
robust performance under cascading node-failure scenarios.
\end{abstract}


\section{Introduction}
\label{sec:introduction}

The rapid proliferation of Earth observation missions has led to an unprecedented growth in remote sensing data generated by Low Earth Orbit (LEO) satellite networks. Traditionally, these systems adopt a \emph{bent-pipe} architecture, where raw data are transmitted to ground stations for centralized processing~\cite{XiaoWC24}. Although effective in early deployments, this paradigm is increasingly constrained by limited downlink bandwidth and the stringent latency requirements of time-sensitive applications such as disaster monitoring and real-time surveillance. As a result, Satellite Edge Computing (SEC) has emerged as a promising alternative, enabling in-orbit data processing and significantly reducing communication overhead. By shifting computation closer to data sources, SEC allows LEO networks to support near real-time intelligent services~\cite{shen2023survey, cao2023edge_assisted}. However, this paradigm shift also fundamentally changes the system operation from centralized processing to distributed, resource-constrained orchestration. In particular, the highly dynamic topology of LEO satellite networks, coupled with limited onboard computation, energy constraints, and intermittent inter-satellite connectivity, leads to severe spatio-temporal resource fragmentation, making efficient and autonomous coordination a critical yet unresolved challenge.

Despite extensive research on routing and resource scheduling in Integrated Satellite-Terrestrial Networks (ISTNs)~\cite{ChenCST25}, existing solutions exhibit fundamental limitations when confronted with the heterogeneous and highly dynamic nature of next-generation LEO satellite networks. 
Most conventional approaches either assume static resource availability or rely on localized optimization frameworks, which systematically fail to capture the global spatio-temporal dynamics inherent to distributed 
SEC environments~\cite{comnet2025multiagent, 
HuangTMC25, PengTNSE26}. 
This limitation becomes particularly pronounced under asymmetric traffic patterns, where large volumes of data generated across the network must be delivered to a small number of geographically constrained ground stations~\cite{lyu2023falcon}. Such many-to-few communication structures inherently create sink-side bottlenecks that cannot be mitigated through local decisions alone. Furthermore, classical heuristic methods lack the adaptability to respond to dynamic disruptions such as topology changes or traffic bursts, while learning-based approaches often suffer from poor generalization due to their dependence on extensive offline training and environment-specific tuning~\cite{comcom2024multiagent, ZhouTMC25}. Consequently, existing solutions exhibit significant performance degradation in dynamic and uncertain environments, highlighting the lack of a unified framework capable of both global reasoning and robust real-time adaptation. This gap underscores the necessity for a new class of cognitive orchestration mechanisms that can leverage prior knowledge while maintaining strong generalization across diverse and evolving network conditions.

To overcome these fundamental limitations, \add{we propose \sysname, a 
\underline{C}ognitive \underline{O}rchestration framework with \underline{R}etrieval-Augmented Generation 
(RAG) \underline{E}nhancement for \underline{LEO} satellite networks.} By shifting the network 
control paradigm from reactive heuristics to knowledge-driven adaptation, 
our framework bridges the gap between high-level cognitive reasoning 
and low-level physical execution. A fidelity-aware genetic scheduler 
guarantees physical feasibility of all dispatched routing and offloading 
decisions, while RAG enables the network to recall Pareto-optimal 
strategies distilled from offline evolutionary optimization, dynamically 
resolving unseen topological bottlenecks in real time. The main 
contributions of this work are outlined as follows:
\begin{itemize}[leftmargin=0.3cm, noitemsep]
    \item We propose an autonomous in-orbit orchestration framework that decouples macroscopic cognitive reasoning from microscopic physical execution, enabling resilient operation without requiring continuous ground support. An abstract state representation compresses distributed telemetry into decision-relevant descriptors, while the bi-level control hierarchy preserves global coordination under asymmetric many-to-few traffic patterns.
    \item We design a bi-level cognitive control mechanism that integrates RAG-grounded strategic reasoning with fidelity-aware genetic execution. The RAG layer anchors Large Language Model (LLM) inference to an Expert Knowledge Base (EKB) of Pareto-optimal historical strategies, preventing hallucination from propagating into the control plane. Building on the inferred preference vector, the genetic scheduler translates high-level decisions into physically admissible routing and computing paths, strictly satisfying resource and topological constraints.
    \item Extensive experiments on a high-fidelity Walker-Delta network testbed demonstrate the effectiveness of CORE-LEO, which achieves a 30.7\% reduction in packet loss, a 30\% improvement in energy efficiency, and an 8.5\% decrease in end-to-end latency against the most competitive learning-based baseline, with robust performance maintained under cascading node-failure scenarios.
\end{itemize}

The remainder of this paper is organized as follows: Section~\ref{sec:rework} summarizes the related work on LEO satellite network orchestration and edge-native satellite networks. Section~\ref{sec:system_model} establishes the system architecture and the bi-level mathematical formulation of the joint routing and computation problem. Section~\ref{sec:methodology} details the proposed framework, elucidating the integration of RAG-driven cognitive inference with the lower-level genetic execution plane. Section~\ref{sec:experiments} provides a comprehensive experimental evaluation, including performance benchmarks and hardware-anchored feasibility analysis. Finally, Section~\ref{sec:conclusion} concludes the work with future research directions.

\section{Related Work}~\label{sec:rework}
Extensive research has explored in-orbit processing, task offloading, and resource allocation to alleviate terrestrial downlink congestion and to reduce service latency in LEO satellite networks~\cite{ChenCST25}. Recent studies emphasize joint optimization strategies, such as integrating computation offloading with power allocation~\cite{xie2025state_delay} or utilizing UAV-assisted SEC configurations~\cite{liu2024sunlight_aware, li2024realtime_satellite_computing, jia2026node_deployment}. Although these studies establish the necessity of collaborative onboard processing, they primarily assume that communication and computation resources are smoothly optimizable or locally manageable. Consequently, these approaches struggle to address the severe spatiotemporal fragmentation of edge computing capacities, intermittent link availability, and strict energy budgets inherent in LEO satellite networks~\cite{rojas2026space_cloud}.

Another research direction investigates contact-aware routing and scheduling in time-varying satellite topologies. Meta-heuristic schedulers, such as Genetic Algorithms (GA), are widely adopted for multi-objective candidate route generation and combinatorial search when exact optimization is intractable~\cite{SOTA_KSP, qi2025ls2moss, zhang2022low_latency_topology, lai2023resilient_routing, li2024skycastle, gu2024starveri}. Despite their utility, most existing solutions rely on static penalty weights or loosely coupled routing-computation paradigms. Under abrupt spatial traffic bursts or cross-plane link failures, these static heuristics fail to dynamically consolidate fragmented resources~\cite{lyu2023falcon, deng2025bottleneck_links, li2025small_scale_leo}. In particular, traditional shortest-path routing often increases network congestion by concentrating traffic on specific nodes, leading to queue buildup even when alternative topological paths are available~\cite{SOTA_KSP, JSAC_UnevenTraffic}.

To enhance network adaptability, Deep Reinforcement Learning (DRL)-based methods have been widely adopted for state-aware resource management~\cite{comnet2025multiagent, comcom2024multiagent, ZhouTMC25, HuangTMC25, tang2024stochastic_offloading, lin2025dependency_aware, rodrigues2023hybrid_learning}. Although these methods outperform handcrafted heuristics by exploiting real-time system states, their efficacy is highly dependent on the specific topology and traffic distributions encountered during training. When subjected to out-of-distribution disruptions, such as cascading node failures or extreme workload heterogeneity, learning-based controllers frequently exhibit degraded robustness and limited generalization capability~\cite{SongCN26}. Therefore, DRL-based approaches face a fundamental trade-off between generalization and training efficiency in highly dynamic non-terrestrial networks.

Recently, LLMs have emerged as cognitive decision-makers for autonomous network management, with RAG providing a critical mechanism for external knowledge grounding~\cite{SOTA_LLM_Net, llm_networking, gao2023rag_survey,shokrnezhad2025arc,huang2026ragsurvey}. Although zero-shot agents and Chain-of-Thought (CoT) prompting demonstrate significant potential in reasoning~\cite{SOTA_CoT, SOTA_ZEROSHOT}, directly deploying LLMs for continuous network orchestration remains perilous. Pure LLM reasoning struggles to enforce strict physical boundaries, such as collision-free transmission windows, dynamic queue evolution, and finite battery safety thresholds. Without an explicit deterministic execution layer, LLM-generated policies risk physical infeasibility, highlighting a profound reasoning-execution gap in resource-constrained environments.

In summary, while prior work has advanced satellite edge offloading, dynamic routing, and learning-based control, three fundamental challenges persist in a unified framework: (i) Inadequate online consolidation of spatio-temporal fragmented resources, (ii) limited generalization under unseen topological anomalies, and (iii) the inability of pure cognitive models to guarantee microscopic physical compliance. To address these limitations, this paper proposes a bi-level orchestration framework that bridges the reasoning-execution gap. By decoupling macroscopic RAG-driven inference from microscopic fidelity-aware execution, we achieve highly adaptive and strictly feasible joint routing and computation scheduling in dynamic SEC networks.

\section{System Model and Problem Formulation}
\label{sec:system_model}

\begin{table}[!t]
\caption{\add{Summary of Key Notations}}
\label{tab:notation}
\centering
\renewcommand{\arraystretch}{1.15}
\begin{tabular}{ll}
\hline\hline
\textbf{Symbol} & \textbf{Description} \\
\hline
\multicolumn{2}{l}{\textit{Network Topology}} \\
$\mathcal{G}(t)$ & Time-dependent network graph \\
$\mathcal{V}, \mathcal{E}(t)$ & Node set and time-varying edge set \\
$\mathcal{S}, \mathcal{G}_{st}$ & Satellite set and ground station set \\
$N, P$ & Number of satellites and orbital planes \\
$\mathcal{W}_{ij}$ & Contact windows of link $(i,j)$ \\
$B_{ij}^{(m)}$ & Bandwidth in the $m$-th contact window \\
\hline
\multicolumn{2}{l}{\textit{Tasks and Resources}} \\
$\mathcal{K}_t$ & Active task batch at epoch $t$ \\
$L_k^{\text{raw}}, L_k'$ & Raw and processed data volume of task $k$ \\
$\rho_k \in (0,1)$ & Compression ratio after in-orbit processing \\
$w_k$ & Computational density (cycles/bit) \\
$C_v^{\text{comp}}$ & CPU capacity of satellite $v$ \\
$E_v^{\max}, E_{\text{safe}}$ & Energy buffer and safety threshold \\
$D_k, t_k^{\text{gen}}$ & Deadline and generation time of task $k$ \\
\hline
\multicolumn{2}{l}{\textit{Communication and Computation}} \\
$\Delta t_{\text{tx}}^k, \Delta t_{\text{prop}}^k$ & Transmission and propagation delay \\
$\Delta t_{\text{proc}}^k$ & Processing duration \\
$\tau_{ij}^k, \tau_v^{\text{proc}}$ & Earliest feasible transmission/processing start \\
$\mathcal{I}_{ij}^{\text{free}}, \mathcal{J}_v^{\text{free}}$ & Residual link/CPU time intervals \\
$T_k^{\text{total}}, D_k^{\text{queue}}$ & End-to-end latency and queueing delay \\
$E_k^{\text{total}}$ & Total energy consumption for task $k$ \\
\hline
\multicolumn{2}{l}{\textit{Decision Variables and Preference}} \\
$\mathcal{P}_k$ & Routing path of task $k$ \\
$x_{k,v} \in \{0,1\}$ & Offloading indicator \\
$\boldsymbol{\omega}_t$ & Preference vector $[\omega_{\text{lat}}, \omega_{\text{cong}}, \omega_{\text{eng}}]^{\top}$ \\
$\mathbf{s}_t, \mathbf{x}_t$ & abstract state descriptor and raw telemetry \\
$\mathcal{H}_t$ & Retrieved historical priors from EKB \\
$\eta_{\text{cpu}}, \eta_{\text{th}}$ & CPU congestion factor and threshold \\
\hline\hline
\end{tabular}
\end{table}

This section describes the physical network models, followed by the communication model, computation model, and energy consumption model. We then present the problem formulation. \add{The key notations used in this paper are summarized in Table~\ref{tab:notation}.}

\subsection{Network Model}
\label{subsec:network_model}

\begin{figure}[t]
    \centering
\includegraphics[width=0.95\linewidth]{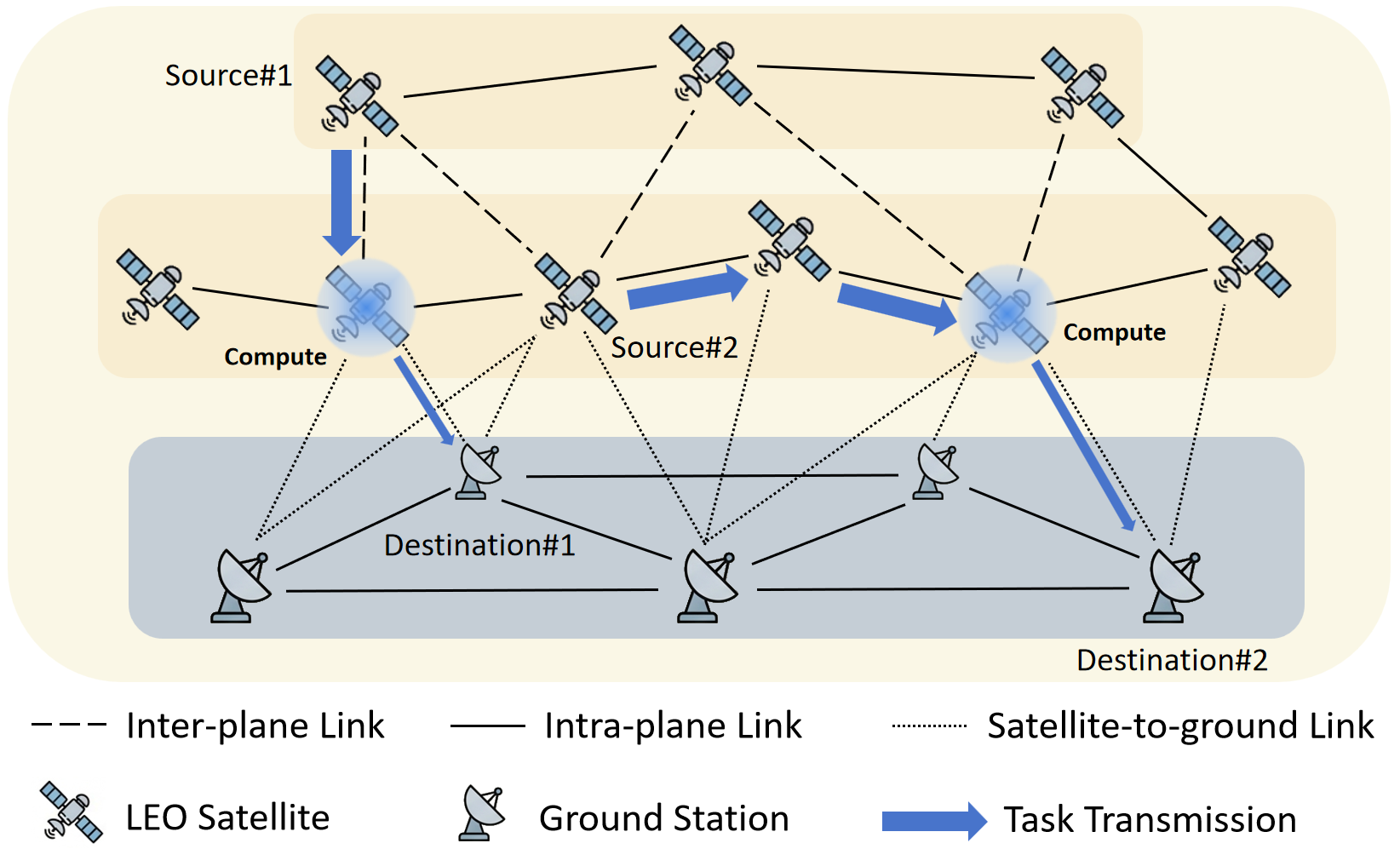}
    \caption{Illustration of the network model.}
    \label{fig:network_model}
\end{figure}

We consider an ISTN aimed at providing connectivity and in-orbit processing capabilities for massive remote sensing tasks. As shown in \figurename~\ref{fig:network_model}, this network operates over a discrete time horizon $T$ corresponding to the network's orbital period.

The physical network topology is inherently time-varying due to the deterministic orbital mechanics of LEO satellites. We model the network as a time-dependent directed graph $\mathcal{G}(t) = (\mathcal{V}, \mathcal{E}(t))$, where the node set $\mathcal{V}$ and edge set $\mathcal{E}(t)$ are defined as follows:

\subsubsection{Heterogeneous Node Set}
The node set $\mathcal{V} = \mathcal{S} \cup \mathcal{G}_{st}$ comprises two distinct types of computational entities:
\begin{itemize}[leftmargin=0.3cm, noitemsep]
    \item \textbf{LEO Satellites ($\mathcal{S}$): }\add{We deploy a constellation of $N$ satellites uniformly distributed across $P$ orbital planes (denoted as an $N \times P$ configuration).} Each satellite $v \in \mathcal{S}$ acts as an in-orbit edge computing node, bounded by finite computational capacity $C_v^{\mathrm{comp}}$ (e.g., CPU cycles per second) and the energy buffer $E_v^{\mathrm{max}}$. 
    \item \textbf{Ground Stations ($\mathcal{G}_{st}$):} A set of globally distributed terrestrial stations acting as data sinks with virtually unlimited power supply compared to satellites.

\end{itemize}

\subsubsection{Heterogeneous Edge Set}
\add{The edge set is composed of four disjoint subsets, i.e.,
\begin{equation}
\mathcal{E}(t) = \mathcal{E}_{\text{intra}}(t) \cup \mathcal{E}_{\text{inter}}(t) 
\cup \mathcal{E}_{\text{sgl}}(t) \cup \mathcal{E}_{\text{ggl}},
\end{equation}
which represents the instantaneous physical connectivity categorized by transmission medium and link stability:}

\begin{itemize}[leftmargin=0.3cm, noitemsep]
   \item \textbf{Inter-Satellite Links (ISLs)}: Comprising stable intra-plane links $\mathcal{E}_{\mathrm{intra}}$ between adjacent nodes and dynamic inter-plane links $\mathcal{E}_{\mathrm{inter}}$ subject to periodic connection loss. \add{Link availability is governed by distance thresholds $D_{\text{intra}}$ and $D_{\text{inter}}$: a link $(i,j)$ is active at time $t$ if and only if $d_{ij}(t) \leq D_{\text{intra}}$ (intra-plane) or $d_{ij}(t) \leq D_{\text{inter}}$ (inter-plane).}

   \item \textbf{Satellite-Ground Links (SGLs):} Opportunistic RF links $\mathcal{E}_{\mathrm{sgl}}$ are established only when the slant range $d_{ij}(t) \le D_{\mathrm{sgl}}$ during a satellite's visibility over a ground station.

   \item \textbf{Ground-Ground Links (GGLs):} High-speed, stable terrestrial fiber backbones $\mathcal{E}_{\mathrm{ggl}}$ connecting ground station $\mathcal{G}_{st}$.

\end{itemize}

\subsubsection{Discrete Contact Window Abstraction}
To resolve the temporal intermittency of the topology, we abstract continuous orbital visibility into a sequence of discrete \textit{contact windows}. For any link $(i, j) \in \mathcal{E}(t)$, its predictable availability over the planning horizon is defined as the set
\begin{equation}
    \mathcal{W}_{ij} = \left\{ t_{ij}^{(m),\mathrm{start}}, t_{ij}^{(m),\mathrm{end}}, B_{ij}^{(m)} \right\}_{m=1}^{M_{ij}},
\label{eq:window}
\end{equation}
where $M_{ij}$ denotes the total number of valid windows and $B_{ij}^{(m)}$ represents the instantaneous bandwidth capacity during the $m$th interval. The capacity is strictly zero outside these intervals, providing the fundamental physical boundary for continuous-time queueing and collision-free task scheduling.

\subsection{Communication Model}
\label{subsec:comm_model}

\add{Unlike terrestrial networks where links are continuously available, LEO satellite network's ISLs and SGLs are only active during disjoint contact windows, causing severe \emph{spatio-temporal resource fragmentation}. Any task transmission must therefore be \emph{scheduled} rather than simply forwarded. To prevent transmission collisions across non-overlapping windows, 
we adopt an event-driven, reservation-based continuous-time queueing model with deterministic admission control: a reservation is accepted only when a contiguous free interval of sufficient duration is available within 
$\mathcal{I}_{ij}^{\text{free}}$.}

\subsubsection{Transmission and Propagation Dynamics}
When task $k$ with data volume $L_k^{\mathrm{current}}$ arrives at node $i$ at time $t_i^{k,\mathrm{arr}}$, 
it is scheduled for forwarding over the directed link $(i, j) \in \mathcal{E}(t)$. $L_k^{\mathrm{current}}$ herein denotes the effective data volume at the current hop, defined as $L_k^{\mathrm{raw}}$ prior to in-orbit processing and $L_k' = \rho_k L_k^{\mathrm{raw}}$ thereafter\add{, where $\rho_k \in (0, 1)$ is the task-specific compression ratio characterizing the data volume reduction induced by in-orbit processing}.

Given the link bandwidth $B_{ij}$ within a valid contact window, the transmission duration is defined as
\begin{equation} \label{eq:tx_duration}
    \Delta t_{\mathrm{tx}}^{k}(i, j) = \frac{L_k^{\mathrm{current}}}{B_{ij}^{(m)}}.
\end{equation}

The propagation delay is distance-dependent, which is formulated as 
\begin{equation}\label{eq:proc_duration}
    \Delta t_{\mathrm{prop}}^{k} = \frac{d_{ij}(\tau_{ij}^k)}{c},
\end{equation}
where $c$ is the speed of light and $\tau_{ij}^k$ \add{denotes the earliest feasible transmission start time on link $(i,j)$}.

\subsubsection{Gap-Finding Queueing Mechanism}
To ensure collision-free scheduling, we define $\mathcal{I}_{ij}^{\mathrm{busy}}$ as the union of all continuous-time intervals already reserved by prior tasks on link $(i, j)$. \add{The set of residual available resources $\mathcal{I}_{ij}^{\text{free}}$  
is obtained as the \emph{set difference} between the physical contact windows 
and the intervals already reserved by prior tasks}:
\begin{equation}  
    \mathcal{I}_{ij}^{\mathrm{free}} = \mathcal{W}_{ij} \setminus \mathcal{I}_{ij}^{\mathrm{busy}}.
\end{equation}

In dynamic ISTNs, traditional First-In-First-Out (FIFO) queueing mechanisms are inadequate due to the periodic \add{connection loss} of ISLs and SGLs. We therefore employ a reservation-based continuous-time queueing model to prevent transmission collisions across disjoint contact windows. \add{Unlike classical TDMA, where time is partitioned into fixed periodic slots assigned \emph{a priori}, our model operates over an \emph{event-driven continuous timeline}: reservations are dynamically inserted into the residual free intervals $\mathcal{I}_{ij}^{\text{free}}$ as tasks arrive, with variable-length slots determined by per-task transmission durations.} Instead of assuming continuous availability, the orchestrator utilizes a \textit{smart gap finding} mechanism. The earliest feasible transmission start time $\tau_{ij}^k$ is determined by identifying the earliest valid intervals within $\mathcal{I}_{ij}^{\mathrm{free}}$ that can accommodate the required duration $\Delta t_{\mathrm{tx}}^{k}$, occurring no earlier than the task's arrival $t_i^{k,\mathrm{arr}}$:
\begin{equation} \label{eq:comm_gap}
    \tau_{ij}^k = \inf \left\{ t \ge t_i^{k,\mathrm{arr}} \mid [t, t + \Delta t_{\mathrm{tx}}^{k}(i, j)] \subseteq \mathcal{I}_{ij}^{\mathrm{free}} \right\}.
\end{equation}

This formulation handles cross-window fragmentation: if a task cannot be completed within the current contact window due to impending orbital occlusion, the operator automatically shifts $\tau_{ij}^k$ to the next available window, inherently capturing the cross-window waiting delay.

\subsubsection{Communication Latency}
The queueing delay at link $(i, j)$ is explicitly tracked as $D_{ij}^{\mathrm{queue}} = \tau_{ij}^k - t_i^{k,\mathrm{arr}}$. Following successful allocation, the arrival time at the downstream node $j$ is updated as
\begin{equation}
    t_j^{k,\mathrm{arr}} = \tau_{ij}^k + \Delta t_{\mathrm{tx}}^{k}(i, j) + \Delta t_{\mathrm{prop}}^{k}.
\end{equation}

This iterative update ensures that the scheduling plan maintains strict temporal causality across the multi-hop satellite path.

\subsection{Computation Model}
\label{subsec:comp_model}

When the orchestrator assigns task $k$ to an intermediate satellite $v \in \mathcal{S}$ for in-orbit processing, the binary variable
$x_{k,v} \in \{0,1\}$ indicates the assignment decision. Specifically, $x_{k,v} = 1$ means task $k$ is assigned to node $v$ for in-orbit processing. The task enters the computational queue of node $v$ immediately upon its complete arrival at absolute time $a_v^k$.

\subsubsection{Processing Duration}
The computational execution duration is determined by the raw data volume $L_k^{\mathrm{raw}}$, the computational workload density $w_k$ (cycles/bit), and the satellite's physical CPU capacity $C_v^{\mathrm{comp}}$ (cycles/s):
\begin{equation}
    \Delta t_{\mathrm{proc}}^{k} = \frac{L_k^{\mathrm{raw}} \cdot w_k}{C_v^{\mathrm{comp}}}.
\end{equation}

\subsubsection{Continuous Computation Gap Finding}
Similar to the communication model, we model the CPU availability of satellite $v$ as a continuous timeline managed by a reservation system. Let $\mathcal{J}_v^{\mathrm{busy}}$ denote the set of time intervals already reserved by preceding tasks on node $v$'s CPU. The residual unreserved CPU time is $\mathcal{J}_v^{\mathrm{free}} = [0, T] \setminus \mathcal{J}_v^{\mathrm{busy}}$.

To avoid computational resource collisions, the start time of the in-orbit processing $\tau_v^{\mathrm{proc}}$ is determined by identifying the earliest contiguous idle CPU gap that can fully accommodate $\Delta t_{\mathrm{proc}}^{k}$ after the data arrival $t_v^{k,\mathrm{arr}}$:
\begin{equation} \label{eq:comp_gap}
    \tau_v^{\mathrm{proc}} = \inf \left\{ t \ge t_v^{k,\mathrm{arr}} \mid [t, t + \Delta t_{\mathrm{proc}}^{k}] \subseteq \mathcal{J}_v^{\mathrm{free}} \right\}.
\end{equation}

\subsubsection{Computation Delay}
The computation queueing delay experienced by task $k$ at node $v$ is defined as $D_v^{\mathrm{queue}} = \tau_v^{\mathrm{proc}} - a_v^k$. Upon completion of the processing phase, the task's data volume is compressed to $L_k' = \rho_k L_k^{\text{raw}}$. The task is then released to the egress communication module at the updated ready time:
\begin{equation}
    t_v^{k,\mathrm{arr}} \leftarrow \tau_v^{\mathrm{proc}} + \Delta t_{\mathrm{proc}}^{k}.
\end{equation}

\subsection{Energy Consumption Model}
\label{subsec:energy_model}

\add{
The energy budget is a critical bottleneck for in-orbit processing. We model 
the per-task energy consumption as the sum of computation and transmission 
expenditures accumulated along the selected multi-hop path $\mathcal{P}_k$.

Let $P_v^{\mathrm{comp}}$ and $P_v^{\mathrm{tx}}$ denote the active 
power consumption coefficients of the CPU and the transceiver at node $v$, respectively. For each node $v \in \mathcal{P}_k$, the per-hop energy contribution comprises a computation term, incurred only if task $k$ is offloaded to $v$ (i.e., $x_{k,v} = 1$), and a transmission term, incurred only if $v$ is not the destination:
\begin{equation} \label{eq:node_energy}
    E_{k,v} = \underbrace{x_{k,v}\, P_v^{\mathrm{comp}}\, \Delta t_{\mathrm{proc}}^{k}}_{\text{computation}}
    + \underbrace{\mathbb{I}(v \neq d_k)\, P_v^{\mathrm{tx}}\, \Delta t_{\mathrm{tx}}^{k}(v, v_{\mathrm{next}})}_{\text{transmission}},
\end{equation}
where $\mathbb{I}(\cdot)$ is the indicator function, $d_k$ is the destination node of task $k$, and $v_{\mathrm{next}}$ denotes the immediate successor of $v$ in $\mathcal{P}_k$. The indicator $\mathbb{I}(v \neq d_k)$ ensures that no transmission energy is charged at the final hop, while 
$\Delta t_{\mathrm{tx}}^{k}$ and $\Delta t_{\mathrm{proc}}^{k}$ have been defined in Eqs.~(\ref{eq:tx_duration}) and~(\ref{eq:proc_duration}), 
respectively.

The total energy expenditure for serving task $k$ is then obtained by summing 
$E_{k,v}$ over all nodes on the routed path:
\begin{equation} \label{eq:energy_total}
    E_k^{\mathrm{total}} = \sum_{v \in \mathcal{P}_k} E_{k,v}.
\end{equation}
}

To ensure the long-term operational survivability of the network, the residual energy $E_v(t)$ of any satellite $v$ must strictly satisfy the safety threshold $E_v(t) \ge E_{\mathrm{safe}}$ at all times, preventing system depletion under extreme compute-bound workloads.

\subsection{Problem Formulation}
\label{subsec:problem_formulation}

The main objective of the proposed orchestration framework is to dynamically resolve spatio-temporal resource fragmentation by balancing end-to-end latency, network resource contention, and global energy consumption. Because this entails both macroscopic semantic trade-offs and microscopic physical scheduling, we formulate the system objective as a bi-level Mixed-Integer Nonlinear Program (MINLP) problem.

\subsubsection{Lower-Level Parameterized Execution}
The execution plane receives a preference vector $\boldsymbol{\omega}_t = [\omega_{\text{lat}}, \omega_{\text{cong}}, \omega_{\text{eng}}]^\top \add{\in \mathbb{R}_{\geq 0}^3}$ for a given task batch $\mathcal{K}_t$ at decision epoch $t$, and computes the discrete routing sequences $\mathcal{P}_k$ and binary offloading decisions $x_{k,v} \in \{0, 1\}$ to minimize the instantaneous weighted fitness cost.

\add{
Let $\mathcal{X}_t = \{ \mathcal{P}_k, x_{k,v} \mid k \in \mathcal{K}_t \}$ denote the joint physical decision space. Given a fixed $\boldsymbol{\omega}_t$, the lower-level scheduler solves:
\begin{align}
    & \operatorname*{minimize}_{\mathcal{X}_t} && J(\mathcal{X}_t \mid \boldsymbol{\omega}_t) = \sum_{k \in \mathcal{K}_t} \Big[ \omega_{\mathrm{lat}} T_k^{\mathrm{total}} + \omega_{\mathrm{cong}} D_k^{\mathrm{queue}} \nonumber \\
    & && \qquad\qquad\qquad + \omega_{\mathrm{eng}} \frac{E_k^{\mathrm{total}}}{E_{\mathrm{ref}}} + \Psi_{\mathrm{deadline}}^k \Big], \label{eq:lower_problem} \\
    & \text{subject to} && \mathcal{P}_k \in \Pi(s_k, d_k \mid \mathcal{G}_t), \quad \forall k \in \mathcal{K}_t, \tag{\the\numexpr\value{equation}+1\relax a} \label{const:topology} \\
    & && \sum_{v \in \mathcal{P}_k} x_{k,v} \leq 1, \quad x_{k,v} \in \{0,1\}, \quad \forall k \in \mathcal{K}_t, \tag{\the\numexpr\value{equation}+1\relax b} \label{const:offload_once} \\
    & && T_k^{\mathrm{total}} = t_{d_k}^{k,\mathrm{arr}} - t_k^{\mathrm{gen}} \leq D_k - t_k^{\mathrm{gen}}, \quad \forall k \in \mathcal{K}_t, \tag{\the\numexpr\value{equation}+1\relax c} \label{const:deadline} \\
    & && [\tau_{ij}^k, \tau_{ij}^k + \Delta t_{\mathrm{tx}}^{k}(i,j)] \subseteq \mathcal{I}_{ij}^{\mathrm{free}}, \quad \forall (i,j) \in \mathcal{P}_k, \tag{\the\numexpr\value{equation}+1\relax d} \label{const:comm_gap} \\
    & && [\tau_v^{\mathrm{proc}}, \tau_v^{\mathrm{proc}} + \Delta t_{\mathrm{proc}}^{k}] \subseteq \mathcal{J}_v^{\mathrm{free}}, \nonumber \\
    & && \qquad \forall v \in \mathcal{P}_k \text{ if } x_{k,v}=1, \tag{\the\numexpr\value{equation}+1\relax e} \label{const:comp_gap} \\
    & && E_v(t) - \sum_{k \in \mathcal{K}_t} E_k^{\mathrm{total}} \geq E_{\mathrm{safe}}, \quad \forall v \in \mathcal{S}. \tag{\the\numexpr\value{equation}+1\relax f} \label{const:energy_safe}
\end{align}
\stepcounter{equation}

Objective~(\ref{eq:lower_problem}) is a preference-weighted aggregation of end-to-end latency $T_k^{\mathrm{total}}$, cumulative queueing delay $D_k^{\mathrm{queue}}$, normalized energy consumption $E_k^{\mathrm{total}}/E_{\mathrm{ref}}$ (with $E_{\mathrm{ref}}$ a normalization constant), and a deadline-violation penalty
\begin{equation} \label{eq:deadline_penalty}
    \Psi_{\mathrm{deadline}}^k = M \cdot \max\!\left(0,\ T_k^{\mathrm{total}} 
    - \left(D_k - t_k^{\mathrm{gen}}\right)\right),
\end{equation}
where $M$ is a large constant enforced during the genetic search to discourage deadline violations. The cumulative queueing delay aggregates per-link and per-node contributions along the routed path:
\begin{equation} \label{eq:queue_delay}
    D_k^{\mathrm{queue}} = \sum_{(i,j) \in \mathcal{P}_k} D_{ij}^{\mathrm{queue}} 
    + \sum_{v \in \mathcal{P}_k} x_{k,v} D_v^{\mathrm{queue}},
\end{equation}
with $D_{ij}^{\mathrm{queue}} = \tau_{ij}^k - t_i^{k,\mathrm{arr}}$ and $D_v^{\mathrm{queue}} = \tau_v^{\mathrm{proc}} - t_v^{k,\mathrm{arr}}$ defined in Sections~\ref{subsec:comm_model} and~\ref{subsec:comp_model}, respectively.

The six constraints (~\ref{const:topology})-(~\ref{const:energy_safe}) of the optimization problem enforce:
\begin{itemize}[leftmargin=0.3cm, noitemsep]
    \item \textbf{Routing feasibility}~(Constraint~(\ref{const:topology})): $\mathcal{P}_k$ must be a valid source-to-destination path on $\mathcal{G}_t$.
    \item \textbf{Single-offload}~(Constraint~(\ref{const:offload_once})): each task is processed in orbit at most once.
    \item \textbf{Hard deadline}~(Constraint~(\ref{const:deadline})): the end-to-end latency is within the latency budget.
    \item \textbf{Collision-free transmission}~(Constraint~(\ref{const:comm_gap})): each reservation fits within the residual free link intervals.
    \item \textbf{Collision-free computation}~(Constraint~(\ref{const:comp_gap})): the processing slot fits within the residual CPU availability.
    \item \textbf{Energy safety}~(Constraint~(\ref{const:energy_safe})): every satellite retains a minimum reserve $E_{\mathrm{safe}}$ at all times.
\end{itemize}
}

\subsubsection{Upper-Level Cognitive Optimization}
Acting as the leader, the cognitive plane seeks an optimal mapping policy to generate preference vectors $\boldsymbol{\omega}_t$ at each decision epoch $t$ that minimize the instantaneous macroscopic system penalty $\mathcal{L}_{\mathrm{macro}}$, \add{e.g., the aggregate packet loss and queue overflow at epoch $t$. Three auxiliary quantities are used in the formulation:
\begin{itemize}[leftmargin=0.3cm, noitemsep]
    \item $\eta_{\mathrm{cpu}}(\mathbf{s}_t) \in \mathbb{R}_{\geq 0}$: the CPU congestion factor, defined as the ratio of the average residual CPU queue time to the theoretical average processing time;
    \item $\eta_{\mathrm{th}} > 0$: a predefined criticality threshold;
    \item $\boldsymbol{\omega}_{\mathrm{safe}} \in \mathbb{R}_{\geq 0}^3$: a conservative preference vector prioritizing energy preservation under critical load.
\end{itemize}
}
The upper-level online decision problem is formulated as
\begin{align}
    & \operatorname*{minimize}_{\boldsymbol{\omega}_t} && \mathcal{L}_{\mathrm{macro}}\big(\mathbf{s}_t, \mathcal{X}^*(\boldsymbol{\omega}_t)\big), \label{eq:upper_level_obj} \\
    & \text{subject to} && \mathcal{X}^*(\boldsymbol{\omega}_t) \in \arg\min_{\mathcal{X}_t} J(\mathcal{X}_t \mid \boldsymbol{\omega}_t), \tag{\the\numexpr\value{equation}+1\relax a} \label{eq:upper_follower} \\
    & && \boldsymbol{\omega}_t = \boldsymbol{\omega}_{\mathrm{safe}}, \quad \text{if } \eta_{\mathrm{cpu}}(\mathbf{s}_t) \geq \eta_{\mathrm{th}}. \tag{\the\numexpr\value{equation}+1\relax b} \label{eq:upper_override}
\end{align}
\refstepcounter{equation}\label{eq:upper_constraints} %
\stepcounter{equation} 

\add{Constraint~(\ref{eq:upper_follower}) encodes the bi-level coupling: 
the upper-level decision $\boldsymbol{\omega}_t$ is evaluated against the 
optimal lower-level response $\mathcal{X}^*(\boldsymbol{\omega}_t)$ defined 
by Objective~(\ref{eq:lower_problem}) and constraintsx(~\ref{const:topology})-(~\ref{const:energy_safe}).}

Taken together, the upper-level problem (\ref{eq:upper_level_obj}) and the lower-level problem (\ref{eq:lower_problem}), \add{including their respective objectives and all feasibility constraints}, constitute a bi-level program that is analytically intractable: the lower-level constraints involve non-differentiable infimum operators ($\inf$) and \add{set-membership conditions} ($\subseteq$) over continuous time, invalidating gradient-based reductions. This intractability motivates the architectural decoupling in Section~\ref{sec:methodology}, where a RAG-augmented LLM generates heuristic preference vectors for the upper level while a fidelity-aware GA handles the lower-level combinatorial search.

\section{Methodology}
\label{sec:methodology}

Building on the architectural decoupling introduced in the previous section, we present a RAG-driven bi-level cognitive orchestration framework for joint routing and computation offloading in LEO satellite networks. The upper layer employs a RAG-augmented large language model to perform context-aware reasoning over the observed network state. Instead of relying on explicit gradient information, it retrieves relevant prior solutions from an offline knowledge base to guide the generation of preference vectors. The lower layer adopts a fidelity-aware GA to explore the feasible solution space and produce physically admissible routing and offloading decisions that are consistent with these preferences. This coordinated design provides a tractable approximation to the original problem while preserving adaptability and execution feasibility under dynamic LEO conditions. The following subsections describe the abstract state abstraction, the preference inference mechanism, and the fidelity-aware scheduling procedure.
\subsection{Overview}
\begin{figure*}[t]
\centering
\includegraphics[width=0.8\linewidth]{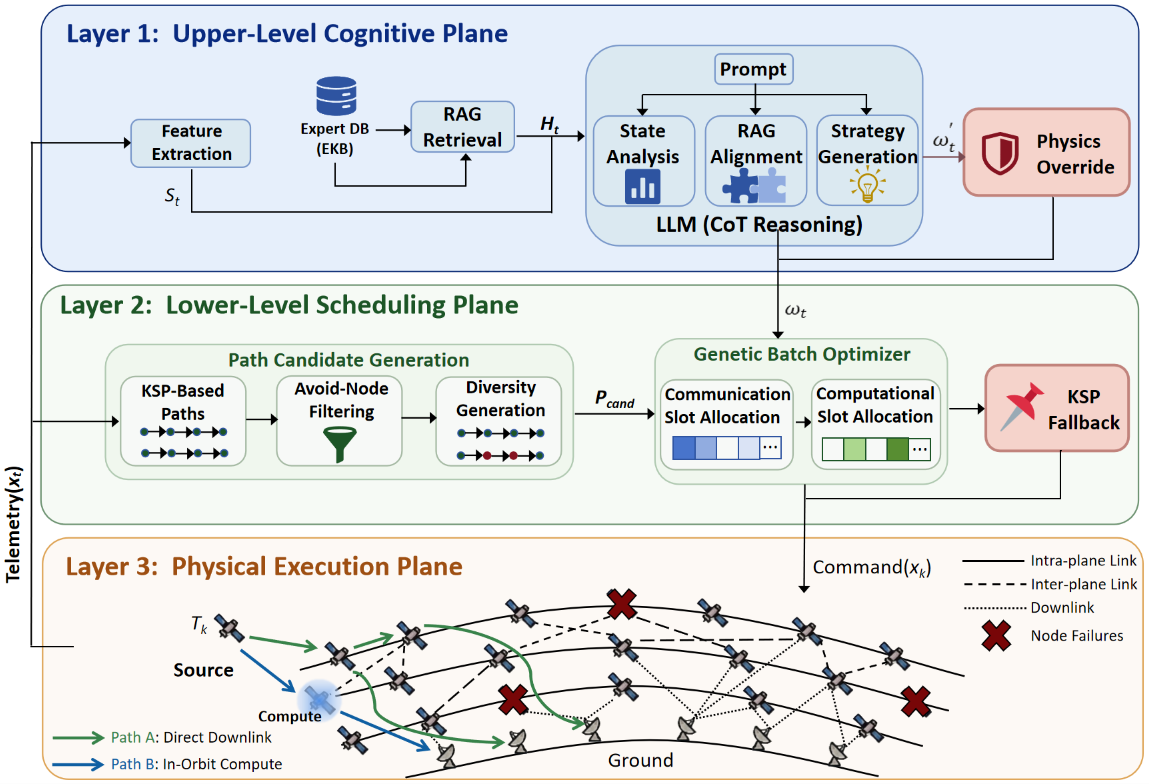}
\caption{\textbf{System Architecture of the Knowledge-Guided Bi-Level Cognitive Orchestration Framework.} The system is structurally decoupled into three interacting planes to bridge context-aware reasoning and physical execution. \textbf{(1) Upper-Level Cognitive Plane:} Extracts compact abstract states ($s_t$) from raw telemetry ($x_t$) and retrieves historical priors ($H_t$) from the EKB. The LLM processes these inputs via Chain-of-Thought reasoning to deduce preference weights. A deterministic physics override mechanism intervenes during severe congestion or LLM hallucination to ensure the final weight ($\omega_t$) is physically safe. \textbf{(2) Lower-Level Scheduling Plane:} Operates as a parameterized solver. It generates state-aware path candidates ($P_{cand}$) using real-time telemetry and employs a Genetic Batch Optimizer, strictly guided by $\omega_t$, to allocate continuous communication and computation slots. A KSP fallback guarantees execution continuity under Genetic Algorithm (GA) timeouts. \textbf{(3) Physical Execution Plane:} The dynamic LEO satellite network that receives and executes the deterministic scheduling commands ($\chi_k$), routing and computing tasks across time-varying topologies and node failures. }
\label{fig:architecture}
\end{figure*}

The cognitive orchestration of LEO satellite networks is formulated in Section~\ref{subsec:problem_formulation} as an intractable bi-level MINLP problem. However, solving this formulation faces two fundamental barriers: the combinatorial complexity of the MINLP renders exhaustive search computationally prohibitive for in-orbit execution, while the 
non-differentiable \add{set-membership conditions} ($\subseteq$) in the continuous-time queueing constraints prevent gradient-based bi-level reductions.

To operationalize this formulation under in-orbit computational constraints, we propose the decomposed bi-level cognitive orchestration framework \sysname as illustrated in \figurename~\ref{fig:architecture}. Rather than deriving an analytical mapping between layers, we operationalize the approximation as a structured inference-execution pipeline. The framework explicitly decouples the bi-level MINLP into two interacting surrogate planes that bridge high-level context-aware reasoning with low-level physical feasibility:

\begin{itemize}[leftmargin=0.3cm, noitemsep]
    \item \textbf{Upper-Level Cognitive Plane (Bi-Level Leader).} To optimize the macroscopic utility in Problem~\eqref{eq:upper_level_obj}--\eqref{eq:upper_constraints} without gradient information from the follower, we employ \textit{retrieval-augmented interpolation}. \add{This plane utilizes an offline-distilled EKB containing Pareto-optimal configurations generated via Evolutionary Multi-Objective Optimization (EMOO). At inference time, a RAG-driven LLM serves as a retrieval-conditioned module: given the abstract state $\mathbf{s}_t$ and the top-$K$ retrieved priors $\mathcal{H}_t$ from the EKB, it produces a preference vector $\boldsymbol{\omega}_t$ that interpolates among the retrieved configurations in a context-aware manner. This sidesteps the analytical intractability of the bi-level problem by replacing gradient-based descent with example-conditioned inference. Boundary safety is enforced via a deterministic physics override mechanism (Eq.~\eqref{eq:upper_override}).}
    
    \item \textbf{Lower-Level Scheduling Plane (Follower Execution):} Considering the cognitive output $\boldsymbol{\omega}_t$ as a definitive exogenous parameter, this plane focuses on the combinatorial minimization defined in Objective~(\ref{eq:lower_problem}). We implement a \textit{fidelity-aware genetic algorithm} as a parameterized combinatorial executor. This component effectively handles the non-differentiable nature of the search space through evolutionary search and a \textit{smart gap-finding} mechanism, which rigidly enforces topological validity (Constraint~(\ref{const:topology})) and collision-free spatio-temporal scheduling (Constraints~(\ref{const:comm_gap}) and (\ref{const:comp_gap})).
\end{itemize}

\subsection{\add{Abstract State Representation}}
\add{High-fidelity satellite telemetry contains redundant and high-frequency 
measurements that, when concatenated across the entire network, form a 
high-dimensional raw observation vector $\mathbf{x}_t$ exceeding the practical context capacity of the reasoning layer. Our framework \sysname therefore applies a 
deterministic feature-extraction mapping at each decision epoch $t$, 
compressing the raw observation into a compact five-dimensional descriptor $\mathbf{s}_t$ that retains only the features relevant to scheduling preference inference.}
 The global raw telemetry at epoch $t$ is collected as
\begin{equation}
\mathbf{x}_t = \left[ \mathbf{q}(t), \mathbf{b}(t), \mathbf{c}(t), \mathbf{e}(t) \right]^\top,
\end{equation}
with the component vectors defined as
\begin{equation}
\begin{aligned}
    \mathbf{q}(t) &= [Q_i(t)]_{i \in \mathcal{V}}, \quad & \mathbf{b}(t) &= [B_{ij}(t)]_{(i,j) \in \mathcal{E}(t)}, \\
    \mathbf{c}(t) &= [C_i(t)]_{i \in \mathcal{V}}, \quad & \mathbf{e}(t) &= [E_i(t)]_{i \in \mathcal{V}},
\end{aligned}
\end{equation}
Here, $Q_i(t)$ denotes the task queue backlog at node $i$, $B_{ij}(t)$ the instantaneous available bandwidth on link $(i,j)$, $C_i(t)$ the computational capacity, and $E_i(t)$ the residual energy.
An abstraction mapping $\Phi: \mathcal{X} \to \mathbb{R}^5$ projects $\mathbf{x}_t$ into a structured descriptor:
\begin{equation}
\mathbf{s}_t =
\left[
\delta_{\mathrm{den}},\;
\bar{\tau}_{\mathrm{ttl}},\;
N_{\mathrm{cong}},\;
N_{\mathrm{fail}},\;
\eta_{\mathrm{cpu}}
\right]^\top,
\end{equation}
where $\delta_{\mathrm{den}}$ is the normalized average data volume per task in the current batch, $\bar{\tau}_{\mathrm{ttl}}$ is the mean deadline window width of incoming tasks, $N_{\mathrm{cong}}$ is the count of inter-satellite links whose maximum predicted queueing delay exceeds a stability threshold, $N_{\mathrm{fail}}$ is the number of satellite nodes currently unavailable due to hardware fault or orbital shielding, and $\eta_{\mathrm{cpu}}$ is the ratio of the average residual CPU queue time to the theoretical average processing time across the network. Each component is computed as follows:
\begin{itemize}[leftmargin=0.3cm, noitemsep]
    \item \textbf{Task Density ($\delta_{\mathrm{den}}$):} It is the normalized average data volume per task in batch $\mathcal{K}_t$:
    \begin{equation}
        \delta_{\mathrm{den}} = \frac{\sum_{k \in \mathcal{K}_t} L_k^{\mathrm{raw}}}{L_{\mathrm{ref}} \cdot |\mathcal{K}_t| + \epsilon},
    \end{equation}
    where $L_k^{raw}$ is the raw data volume of task $k$, $L_{\mathrm{ref}}$ is a reference data volume for normalization, and $\epsilon$ prevents zero-division.

    \item \textbf{Average Time-To-Live ($\bar{\tau}_{\mathrm{ttl}}$):} The mean deadline window of tasks in $\mathcal{K}_t$, measured from task generation time:
    \begin{equation}
        \bar{\tau}_{\mathrm{ttl}} = \frac{1}{|\mathcal{K}_t|} \sum_{k \in \mathcal{K}_t} \left( D_k - t_k^{\mathrm{gen}} \right),
    \end{equation}
    where $D_k$ and $t_k^{\mathrm{gen}}$ are the absolute deadline and generation time of task $k$, respectively.

    \item \textbf{\add{Number of Congested Links} ($N_{\mathrm{cong}}$):} The number of inter-satellite links whose maximum predicted queueing delay exceeds threshold $T_{\mathrm{cong}}$, computed from the occupied slot set $\mathcal{I}_{ij}^{\mathrm{busy}}$ (the complement of the free slot set $\mathcal{I}_{ij}^{\mathrm{free}}$ defined in Section~\ref{subsec:comm_model}):
    \begin{equation}
        N_{\mathrm{cong}} = \sum_{(i,j) \in \mathcal{E}(t)} \mathbb{I} \left( \max_{\add{\tau} \in \mathcal{I}_{ij}^{\mathrm{busy}}} (\add{\tau}^{\mathrm{end}} - t) > T_{\mathrm{cong}} \right),
    \end{equation}
    
    where $\tau^{\mathrm{end}}$ is the end time of a scheduled slot, corresponding to approximately one inter-satellite handover interval.

    \item \textbf{\add{Number of Failed Nodes} ($N_{\mathrm{fail}}$):} The number of satellite nodes currently unavailable due to hardware fault or orbital shielding:
    \begin{equation}
        N_{\mathrm{fail}} = \sum_{v \in \mathcal{S}} \mathbb{I}\!\left(\mathrm{status}(v) = \textsc{Failed}\right).
    \end{equation}

    \item \textbf{CPU Congestion Factor ($\eta_{\mathrm{cpu}}$):} The ratio of the average residual CPU queue time to the theoretical average processing time across the network:
    \begin{equation}
        \eta_{\mathrm{cpu}} = \frac{\dfrac{1}{|\mathcal{S}|} \displaystyle\sum_{v \in \mathcal{S}} \max\!\left(0,\, q_v^{\mathrm{end}} - t\right)}{\bar{L} \cdot \bar{w} / \bar{C}},
    \end{equation}
    where  $|\mathcal{S}|$ denotes the total number of satellites, $q_v^{\mathrm{end}}$ is the completion time of the last queued task at node $v$, and $\bar{L}$, $\bar{w}$, $\bar{C}$ are the epoch-wise network averages of data size, workload density, and computational capacity, respectively.
\end{itemize}

The resulting descriptor $\mathbf{s}_t$ retains the scheduling-relevant features of $\mathbf{x}_t$ while reducing the input volume to a fixed-length representation suitable for prompt construction.

\subsection{Upper-Level Cognitive Plane: RAG-Driven Preference Inference}~\label{subsec:ul}
Within the proposed bi-level architecture, the upper-level cognitive plane acts as the strategic leader. Its primary function is to dynamically determine the global preference vector, which subsequently serves as the exogenous parameter governing the lower-level execution objective.

We define the continuous preference vector at time $t$ as
\begin{equation}
\boldsymbol{\omega}_t =
[\omega_{\mathrm{lat}}, \omega_{\mathrm{cong}}, \omega_{\mathrm{eng}}]^{\top},
\end{equation}
\add{whose entries balance the relative importance of end-to-end latency, 
congestion delay, and energy consumption, respectively, in the lower-level 
Objective~(\ref{eq:lower_problem}).}

The  objective of the upper-level leader is to determine the optimal preference vector that minimizes the expected global scheduling cost across varying network states:
\begin{equation}
\operatorname*{minimize}_{\boldsymbol{\omega}_t} \quad 
\mathbb{E}_{\mathbf{s}_t} 
\left[ J\left( \mathcal{X}^*(\mathbf{s}_t, \boldsymbol{\omega}_t) \right) \right],
\label{eq:upperlevel}
\end{equation}
where $\mathcal{X}^*(\mathbf{s}_t,\boldsymbol{\omega}_t)$ denotes the optimal physical scheduling plan generated by the lower-level follower layer.

Solving Eq.~(\ref{eq:upperlevel}) directly via analytical methods is computationally intractable. Each evaluation of the outer objective requires the exact resolution of a combinatorial scheduling problem. To approximate this optimization \add{within a bounded computation budget compatible with the in-orbit decision epoch} without relying on lower-level gradients, we substitute analytical derivation with a RAG-driven inference pipeline, anchored by an offline-distilled EKB.

\textbf{Offline Knowledge Distillation}: To establish a rigorous ground truth for the cognitive layer, representative extreme network states are generated via large-scale Monte Carlo traffic simulations. For each sampled \add{abstract} state $\mathbf{s}_{\mathrm{off}}$, the Non-dominated Sorting Genetic Algorithm II (NSGA-II)\add{~\cite{deb2002nsga2}} is used to compute the Pareto-optimal scheduling frontier. Each resulting strategy yields a state-preference pair $\langle \mathbf{s}_\text{off}, \boldsymbol{\omega}_\text{off} \rangle$, which is cataloged within the EKB as a verified physical prior.

\textbf{Online Preference Generation:} Let $E(\cdot)$ denote the embedding function that maps an abstract state into a high-dimensional vector representation. At runtime, the system queries the EKB using the current abstract state $\mathbf{s}_t$. To isolate the most relevant historical priors, the retrieval mechanism first applies a metadata filter (bounding load factors and fault counts) to define a physically comparable subspace $\mathcal{D}_t \subseteq \mathtt{EKB}$. Within this constrained subspace, the top-$K$ retrieval is formalized as finding the subset $\mathcal{H}_t$ of size $K$ that maximizes the cumulative cosine similarity:
\begin{equation}
\mathcal{H}_t = \underset{\mathcal{H} \subseteq \mathcal{D}_t, \, |\mathcal{H}| = K}{\arg\max} \sum_{\langle \mathbf{s}_\text{off}, \boldsymbol{\omega}_\text{off} \rangle \in \mathcal{H}} \mathrm{sim}(\mathbf{s}_t, \mathbf{s}_\text{off}).
\end{equation}

\add{Cosine similarity is adopted over Euclidean distance for two reasons: 
(i) it is invariant to the absolute magnitudes of the descriptor entries, 
making the retrieval robust across heterogeneous network scales, and (ii) 
the cumulative formulation favors candidate sets that are collectively 
close to $\mathbf{s}_t$ along multiple semantic directions rather than 
clustered around a single nearest prior, which improves the diversity of 
priors supplied to the downstream LLM for in-context interpolation.}
where the similarity between the current and historical abstract states is evaluated as
\begin{equation}
\mathrm{sim}(\mathbf{s}_t, \mathbf{s}_\text{off}) = \frac{E(\mathbf{s}_t)^\top E(\mathbf{s}_\text{off})}{\|E(\mathbf{s}_t)\|_2 \|E(\mathbf{s}_\text{off})\|_2}.
\end{equation}

Subsequently, a local LLM functions as a preference parameter generator. By processing the retrieved context $\mathcal{H}_t$ via a structured CoT prompt, the LLM semantically interpolates between the historical priors to infer the adaptive preference vector:
\begin{equation}
\boldsymbol{\omega}_t =
\mathrm{LLM}(\mathbf{s}_t,\mathcal{H}_t),
\end{equation}
where $\mathrm{LLM}$ represents the frozen model parameters.

\textbf{Deterministic Physics Override}: \add{Although the LLM is conditioned 
upon the retrieved priors and the abstract state, its output $\boldsymbol{\omega}_t'$ remains a learned mapping with no formal guarantee of physical safety: under unseen state regimes, the inferred preference may deviate from operationally viable configurations and drive the network into unstable states. To prevent such anomalous outputs from propagating into the execution plane, the inference pipeline incorporates a deterministic physical guardrail.} When severe resource bottlenecks are detected, i.e., the CPU congestion factor exceeds a critical threshold ($\eta_{\mathrm{cpu}} \geq \eta_{\mathrm{th}}$), the LLM inference is bypassed and a conservative preference vector $\boldsymbol{\omega}_{\mathrm{safe}}$ is enforced to prevent computational overload.

The detailed inference procedure is implemented as a retrieval–generation pipeline with deterministic guardrails, as summarized in Algorithm~\ref{alg:rag_inference}, where $\tau_{\mathrm{hard}}$ is a hard similarity threshold for retrieved priors, 
and $\Omega_{\mathrm{bounds}}$ defines the feasible range of preference weights.

\begin{algorithm}[t!]
\caption{RAG-Driven Cognitive Preference Inference}
\label{alg:rag_inference}
\begin{algorithmic}[1]
\REQUIRE System state $\mathbf{s}_t$, Expert Knowledge Base $\textit{EKB}$, congestion ratio $\eta_{\mathrm{cpu}}$, threshold $\eta_{\mathrm{th}}$
\ENSURE Bounded penalty weights $\boldsymbol{\omega}_t$
\STATE $q_t \leftarrow \text{GenerateFingerprint}(\mathbf{s}_t)$
\STATE $\mathcal{H}_t \leftarrow \text{Retrieve}(\textit{EKB}, q_t, \text{Top-}K)$
\STATE $\mathcal{H}_t^{*} \leftarrow \text{Filter}(\mathcal{H}_t, \tau_{\mathrm{hard}})$
\IF{$\mathcal{H}_t^{*} \neq \emptyset$}
    \STATE context $\leftarrow \text{BuildPrompt}(\mathcal{H}_t^{*})$
\ELSE
    \STATE context $\leftarrow \text{BuildPrompt}(\mathcal{H}_t)$
\ENDIF
\STATE $\boldsymbol{\omega}_t \leftarrow \text{Parse}(\text{LLM}(q_t, \text{context}))$
\IF{$\eta_{\mathrm{cpu}} \ge \eta_{\mathrm{th}}$}
    \STATE $\boldsymbol{\omega}_t \leftarrow \text{PhysicsOverride}(\boldsymbol{\omega}_t)$
\ENDIF
\STATE $\boldsymbol{\omega}_t \leftarrow \text{Clip}(\boldsymbol{\omega}_t, \Omega_{\mathrm{bounds}})$
\RETURN $\boldsymbol{\omega}_t$
\end{algorithmic}
\end{algorithm}

\subsection{Lower-Level Scheduling Plane: Fidelity-Aware Execution}
\add{The lower-level scheduling problem, formulated in 
Section~\ref{subsec:problem_formulation} as an MINLP comprising 
Objective~(\ref{eq:lower_problem}) subject to 
constraints~(\ref{const:topology})--(\ref{const:energy_safe}), 
is defined over a combinatorial path-and-offloading space whose feasibility 
hinges on non-differentiable set-membership conditions over the residual 
free intervals $\mathcal{I}_{ij}^{\mathrm{free}}$ and $\mathcal{J}_v^{\mathrm{free}}$. 
The non-linearity precludes conventional Mixed-Integer Linear Programming 
(MILP) solvers, while the non-differentiable set-membership constraints 
rule out gradient-based methods, as constraint satisfaction requires a 
custom continuous-time gap-finding evaluator rather than algebraic 
feasibility checks. We therefore adopt the GA as the 
lower-level solver, which 
(i) accommodates non-differentiable fitness evaluation via gap-finding, 
(ii) supports multi-objective combinatorial search through population-based 
exploration, and (iii) operates under a bounded time budget aligned with 
the in-orbit decision epoch.}

Operating as the follower in the bi-level hierarchy, the lower-level 
scheduling plane receives the preference vector $\boldsymbol{\omega}_t$ 
from the upper-level cognitive plane. Its objective is to determine 
physically feasible routing paths and computation offloading decisions 
for the active task batch $\mathcal{K}_t$. For each task 
$k \in \mathcal{K}_t$, the scheduling decision is formalized by the tuple 
$\chi_k = (\mathcal{P}_k, x_{k,v})$. The path component 
$\mathcal{P}_k \in \mathcal{P}_{\mathrm{cand}}$ denotes the selected routing 
path from the candidate path set, and the binary indicator 
$x_{k,v} \in \{0,1\}$ designates the offloading status of task $k$ on 
satellite edge node $v$. \add{The single in-orbit compression requirement 
$\sum_{v \in \mathcal{P}_k} x_{k,v} \le 1,\ \forall k \in \mathcal{K}_t$ 
is inherited from constraint~(\ref{const:offload_once}).}

To ensure execution fidelity under the highly dynamic topologies of LEO satellite networks, the generated schedules must adhere to the physical constraints derived in Section~\ref{sec:system_model}.

Communication feasibility is enforced by the gap-finding constraint Eq.~(\ref{const:comm_gap}), which ensures that the transmission is scheduled within active contact windows with sufficient bandwidth capacity.For computation, node $v$'s processing feasibility is enforced through 
the continuous-time gap-finding mechanism Eq.~(\ref{eq:comp_gap}), 
which prevents concurrent task execution on the same CPU, 
ensuring each task is allocated a non-overlapping time slot 
within the node's available capacity $C_v^{\mathrm{comp}}$.

For computation, the feasibility of node $v$'s processing capacity $C_v^{\mathrm{comp}}$ is enforced through the continuous-time gap-finding mechanism (Eq.~(\ref{eq:comp_gap})), which prevents overlapping task executions and thereby ensures that instantaneous workload at any node never exceeds its capacity.
The lower-level optimization is solved by two specialized algorithmic components detailed in Algorithm~\ref{alg:path} and Algorithm~\ref{alg:schedule}: a hybrid routing candidate generator and a fidelity-aware genetic scheduler that jointly search the combinatorial space under the guidance of $\boldsymbol{\omega}_t$.

\begin{algorithm}[t!]
\caption{Hybrid Path Candidate Generation}
\label{alg:path}
\begin{algorithmic}[1]
\REQUIRE Active graph $\mathcal{G}_t=(\mathcal{V}, \mathcal{E}_t)$, source $s$, destination $d$, path count $K$, perturbation range $\delta$, inflation factor $\alpha$
\ENSURE Candidate path set $\mathcal{P}_{\mathrm{cand}}$
\STATE $\mathcal{P}_{\mathrm{cand}} \leftarrow \text{K-ShortestPath}(\mathcal{G}_t, s, d)$
\STATE $\mathcal{G}'_t \leftarrow \mathcal{G}_t$
\FOR{each $e \in \mathcal{E}_t$ sorted by $D_e^{\mathrm{queue}}$ descending}
    \STATE $w_e \leftarrow w_e \times \alpha$ \COMMENT{Update weight in $\mathcal{G}'_t$}
    \STATE $\mathcal{P}_{\mathrm{cand}} \leftarrow \mathcal{P}_{\mathrm{cand}} \cup \{\text{ShortestPath}(\mathcal{G}'_t, s, d)\}$
\ENDFOR
\WHILE{$|\mathcal{P}_{\mathrm{cand}}| < K$}
    \FOR{each $e \in \mathcal{G}_t$}
        \STATE $w_e \leftarrow w_e \times \mathrm{Uniform}(1-\delta, 1+\delta)$
    \ENDFOR
    \STATE $\mathcal{P}_{\mathrm{cand}} \leftarrow \mathcal{P}_{\mathrm{cand}} \cup \{\text{ShortestPath}(\mathcal{G}_t, s, d)\}$
\ENDWHILE
\RETURN Top-$K$ unique valid paths in $\mathcal{P}_{\mathrm{cand}}$
\end{algorithmic}
\end{algorithm}

\subsection{Closed-Loop Cognitive Orchestration Workflow}

To bridge the context-aware reasoning of the cognitive leader and the combinatorial execution of the scheduling follower, we formulate the overall system as a bi-level, closed-loop orchestration process. Rather than treating preference generation and scheduling
as independent modules, \sysname integrates preference generation and schedule execution into a unified event-triggered control loop. In this paradigm, high-level inference dynamically parameterizes the lower-level execution only when statistically significant network anomalies occur, thereby conserving vital in-orbit computational resources.

At each decision epoch $T_D$, the system evaluates the abstract state $\mathbf{s}_t$. If an anomaly is detected (e.g., topological failure $N_{\mathrm{fail}} > 0$ or severe congestion $N_{\mathrm{cong}}$ exceeds historical bounds), the RAG-driven cognitive plane is invoked to deduce an adaptive preference vector $\boldsymbol{\omega}_t$, which subsequently guides the evolutionary scheduler. Conversely, under normal network conditions, the computationally expensive inference is bypassed.

To guarantee task servicing and system stability, our framework incorporates a deterministic heuristic fallback. Any tasks left unscheduled by the evolutionary algorithm due to timeout constraints or bypassed epochs (denoted as the unallocated set $\mathcal{U}_t$) are immediately routed using a low-overhead $K$-Shortest Path (KSP) baseline. Furthermore, the closed-loop process includes a self-evolution mechanism: highly successful anomaly-resolution strategies evaluated via empirical physical feedback are iteratively appended to the $\textit{EKB}$, continuously enriching the EKB with verified scheduling priors.

The complete event-triggered orchestration workflow is detailed in Algorithm~\ref{alg:orchestration}, where $N_{\min}$ is the minimum batch size required for a strategy to be 
considered statistically representative,  $\tau_{\mathrm{success}}$ is 
the minimum empirical performance score for EKB admission, and $f_t$ is the empirical scheduling performance score, 
e.g., the fraction of tasks meeting their deadlines in epoch $t$.

\begin{algorithm}[t!]
\caption{Bi-Level Cognitive Orchestration}
\label{alg:orchestration}
\begin{algorithmic}[1]
\REQUIRE Task batch $\mathcal{K}_t$, Knowledge Base $\textit{EKB}$, decision interval $T_D$
\ENSURE Execution plan $\mathcal{X}_t$
\FOR{each decision epoch $t$ with step $T_D$}
    \STATE $\mathbf{x}_t \leftarrow \text{ObserveState}()$
    \STATE $\mathbf{s}_t \leftarrow \Phi(\mathbf{x}_t)$
    \STATE Initialize execution plan $\mathcal{X}_t \leftarrow \emptyset$
    \IF{$\text{AnomalyDetected}(\mathbf{s}_t)$}
        \STATE $\boldsymbol{\omega}_t \leftarrow \text{Algorithm~\ref{alg:rag_inference}}(\mathbf{s}_t, \textit{EKB})$
        \STATE $\mathcal{P}_{\mathrm{cand}} \leftarrow \text{Algorithm~\ref{alg:path}}(\mathcal{G}_t)$
        \STATE $\mathcal{X}_t^{GA} \leftarrow \text{Algorithm~\ref{alg:schedule}}(\mathcal{K}_t, \mathcal{P}_{\mathrm{cand}}, \boldsymbol{\omega}_t)$
    \ELSE
        \STATE $\mathcal{X}_t^{GA} \leftarrow \emptyset$ \COMMENT{Normal condition: skip GA, fall back to KSP}
    \ENDIF
    \STATE $\mathcal{U}_t \leftarrow \mathcal{K}_t \setminus \text{Scheduled}(\mathcal{X}_t^{GA})$ \COMMENT{Identify unscheduled tasks}
    \STATE $\mathcal{X}_t^{KSP} \leftarrow \text{KSP\_Baseline}(\mathcal{U}_t, \mathbf{s}_t)$
    \STATE $\mathcal{X}_t \leftarrow \mathcal{X}_t^{GA} \cup \mathcal{X}_t^{KSP}$
    \STATE Execute $\mathcal{X}_t$ and observe feedback $f_t$
    \IF{$|\mathcal{K}_t| \ge N_{\min}$ \AND $f_t \ge \tau_{\mathrm{success}}$}
        \STATE $\textit{EKB} \leftarrow \textit{EKB} \cup \{(\mathbf{s}_t, \boldsymbol{\omega}_t, f_t)\}$
    \ENDIF
\ENDFOR
\end{algorithmic}
\end{algorithm}

To operate the lower-level combinatorial optimization formulated in Objective~\ref{eq:lower_problem}, the framework relies on specialized algorithmic solvers. These solvers, explicitly implemented within the core orchestration loop (Algorithm~\ref{alg:orchestration}, lines 6-7), function as the physical execution engine. They consist of a hybrid routing candidate generator, which explores topologically viable pathways, and a fidelity-aware genetic scheduler, which searches the continuous-time resource allocation space guided by the exogenous preference vector $\boldsymbol{\omega}_t$.

\begin{algorithm}[t!]
\caption{Fidelity-Aware Genetic Scheduling}
\label{alg:schedule}
\begin{algorithmic}[1]
\REQUIRE Task batch $\mathcal{K}_t$, candidate paths $\mathcal{P}_{\mathrm{cand}}$, cognitive preference $\boldsymbol{\omega}_t$
\ENSURE Best feasible schedule $\mathcal{X}^*$
\STATE Initialize population $\mathbb{P}_0$ of size $N_{\mathrm{pop}}$ via random selections from $\mathcal{P}_{\mathrm{cand}}$
\FOR{generation $g = 1$ to $G$}
    \FOR{each chromosome $\mathcal{X}_i \in \mathbb{P}_g$}
        \STATE $\tau_{\mathrm{tx}},\, \tau_{\mathrm{comp}} \leftarrow \mathrm{GapFinding}(\mathcal{X}_i,\, \{\mathcal{W}_{ij}\},\, \{\mathcal{I}_{ij}^{\mathrm{free}}\})$
        \IF{allocation is collision-free \AND satisfies node energy bounds}
            \STATE Compute $T_k^{\mathrm{total}},\, D_k^{\mathrm{queue}},\, E_k^{\mathrm{total}}$ and $\Psi_{\mathrm{deadline}}^k$ for each $k \in \mathcal{K}_t$
            \STATE Calculate $J(\mathbf{X}_i \mid \boldsymbol{\omega}_t)$ based on Eq.~(\ref{eq:lower_problem}).
        \ELSE
            \STATE $J(\mathbf{X}_i \mid \boldsymbol{\omega}_t) \leftarrow \infty$ \COMMENT{Penalize infeasible schedules}
        \ENDIF
    \ENDFOR
    \STATE $\mathbb{P}_{\mathrm{elite}} \leftarrow \mathrm{ExtractTopSchedules}(\mathbb{P}_g)$
    \STATE $\mathbb{P}_{g+1} \leftarrow \mathrm{TournamentSelection}(\mathbb{P}_g, f)$
    \STATE Apply $\mathrm{Crossover}(p_c)$ and $\mathrm{Mutation}(p_m)$ to $\mathbb{P}_{g+1}$
    \STATE $\mathbb{P}_{g+1} \leftarrow \mathrm{Merge}(\mathbb{P}_{g+1},\, \mathbb{P}_{\mathrm{elite}})$
\ENDFOR
\RETURN $\mathcal{X}^* \leftarrow \arg\min_{\mathcal{X} \in \mathbb{P}_G} J(\mathcal{X} \mid \boldsymbol{\omega}_t)$
\end{algorithmic}
\end{algorithm}

\subsection{Computational Complexity Analysis}
The computational complexity of the proposed framework consists of three main components:
\begin{itemize}[leftmargin=0.3cm, noitemsep]
\item \textbf{Preference Inference}: Retrieving Top-$K$ similar states from a database of size $N$ requires $O(Nd)$ time, where $d = 5$ is the dimension of the \add{abstract} state vector. The subsequent LLM inference introduces an overhead of $O(L_{\mathrm{ctx}}^2)$\add{, where $L_{\mathrm{ctx}}$ is the prompt context length. Since our prompt template (state descriptor, top-$K$ retrieved priors, CoT instructions) has a deterministic upper bound on length independent of the network size, $L_{\mathrm{ctx}}$ is bounded by a constant, and the inference overhead reduces to $O(1)$ asymptotically}.

    \item \textbf{Path Generation:} Algorithm~\ref{alg:path} incurs $\mathcal{O}\bigl((|\mathcal{E}|+K)(|\mathcal{E}|+|\mathcal{V}|)\log|\mathcal{V}|\bigr)$ via repeated Dijkstra searches, where the first term accounts for the congestion-aware weight inflation phase and the second for the stochastic perturbation phase.

    \item \textbf{Genetic Scheduling}: Algorithm~\ref{alg:schedule} evaluates scheduling plans across $G$ generations with population size $P$. Since each chromosome evaluation invokes the gap-finding mechanism over up to $|\mathcal{K}_t|$ previously reserved slots per hop, the per-chromosome cost is $O(|\mathcal{K}_t|^2 \cdot L)$, where $L$ denotes the maximum path length. 
\add{The complexity of the genetic scheduling step alone is therefore} 
$O(G \cdot P \cdot |\mathcal{K}_t|^2 \cdot L)$.
\end{itemize}

\add{Combining the three components, the overall per-decision-epoch complexity is $\mathcal{O}\big( Nd + 1 + (|\mathcal{E}|+K)(|\mathcal{E}|+|\mathcal{V}|)\log|\mathcal{V}| + G \cdot P \cdot |\mathcal{K}_t|^2 \cdot L \big)$,} \add{where the four terms correspond to retrieval, LLM inference, path 
generation, and genetic scheduling, respectively.}

\section{Performance Evaluation}~\label{sec:experiments}
This section presents a comprehensive performance evaluation to validate the efficacy, robustness, and scalability of the proposed bi-level cognitive orchestration framework. We first detail the high-fidelity simulation environment. Subsequently, the proposed \sysname is validated against state-of-the-art baselines  under varying traffic densities and extreme topological anomalies. Finally, we analyze the execution overhead to confirm the real-time operational feasibility of the event-triggered cognitive inference mechanism.

\subsection{Simulation Setup}
\label{subsec:setup}
We evaluate the effectiveness of \sysname using our developed LEO satellite network simulation testbed. The evaluation covers three aspects: aggregate scheduling performance under varying traffic loads, resilience under progressive fault injection, and computational feasibility on a reference onboard platform. 

\subsubsection{Network Architecture}
The network is modeled as a Walker Delta 36/3/1 configuration ($N=36$ satellites, $P=3$ orbital planes) at a quasi-polar inclination of $86.4^\circ$ and an altitude of $780$\:km. Real-time orbital dynamics and visibility are propagated via the \texttt{Skyfield} astrodynamics engine utilizing the SGP4 model{\footnote{\url{https://rhodesmill.org/skyfield/}}}. \add{At an altitude of $780$\:km, the resulting orbital period is $T \approx 5400$\,s, as determined by Kepler's third law for LEO.} The network topology comprises three distinct link types:
\begin{itemize}[leftmargin=0.3cm, noitemsep]
    \item \textbf{ISLs}: Intra-plane links are stable links between adjacent satellites in the same plane ($d \le 5000$\:km, $50$\:Gbps), while inter-plane links are dynamic laser links across adjacent planes ($d \le 4500$\:km, $50$\:Gbps), subject to periodic handovers.
    \item \textbf{SGLs}: Opportunistic RF links to $9$ globally distributed ground stations (e.g., Svalbard, Tokyo, Santiago) with a $3000$\:km slant range and $3$\:Gbps capacity.
    \item \textbf{GGLs}: Static $100$\:Gbps terrestrial fiber backbones interconnecting ground sinks.

\end{itemize}

Each satellite node is constrained by a $100$\:GFLOPS computational capacity and a total energy buffer of $5$\:MJ. The power model assumes a $200$\:W peak for computational loads and $40$\:W for each active transceiver link.

\subsubsection{Traffic Generation and Workload Heterogeneity}

Motivated by the documented spatiotemporal traffic imbalances in real-world LEO 
satellite networks~\cite{JSAC_UnevenTraffic}, we 
synthesize a heterogeneous workload comprising two superimposed components: 
\textbf{Continuous Background Telemetry} following a Poisson arrival process, and \textbf{Spatiotemporal Traffic Bursts} injecting concurrent tasks toward specific ground stations to simulate spatial bottlenecks.

\add{To capture the heterogeneous and multi-modal characteristics typical 
of Earth Observation data streams~\cite{lyu2023falcon}, 
tasks are classified into three operational profiles:}
\begin{itemize}[leftmargin=0.3cm, noitemsep]
\item \textbf{Raw Sensor Data} ($p=1$): Large data blocks with low 
computational demand, \add{lowest} priority, and minimal compression ratio.
\item \textbf{Standard Processed Data} ($p=3$): Intermediate payloads with 
moderate computational demand, \add{intermediate} priority, and moderate 
compression ratio.

\item \textbf{Mission-Critical Data} ($p=5$): Compute-intensive tasks with 
high CPU demand, \add{highest} priority, and high compression ratio.
\end{itemize}

\begin{figure}[!t]
    \centering
    \subfloat[Data size distribution\label{fig:task_dist:a}]{%
        \includegraphics[width=0.48\linewidth]{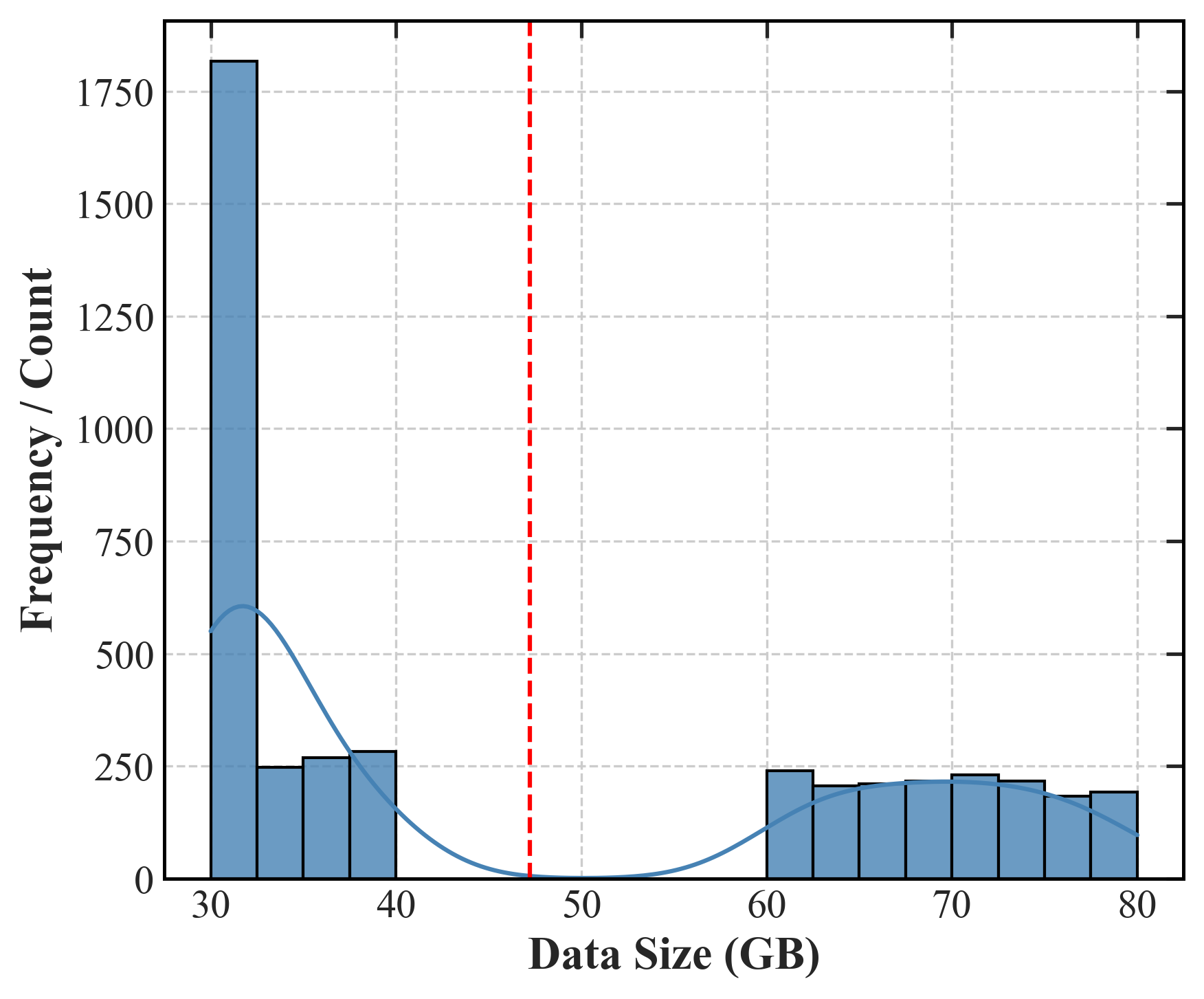}%
    }
    \hfil 
    \subfloat[Computational workload density\label{fig:task_dist:b}]{%
        \includegraphics[width=0.48\linewidth]{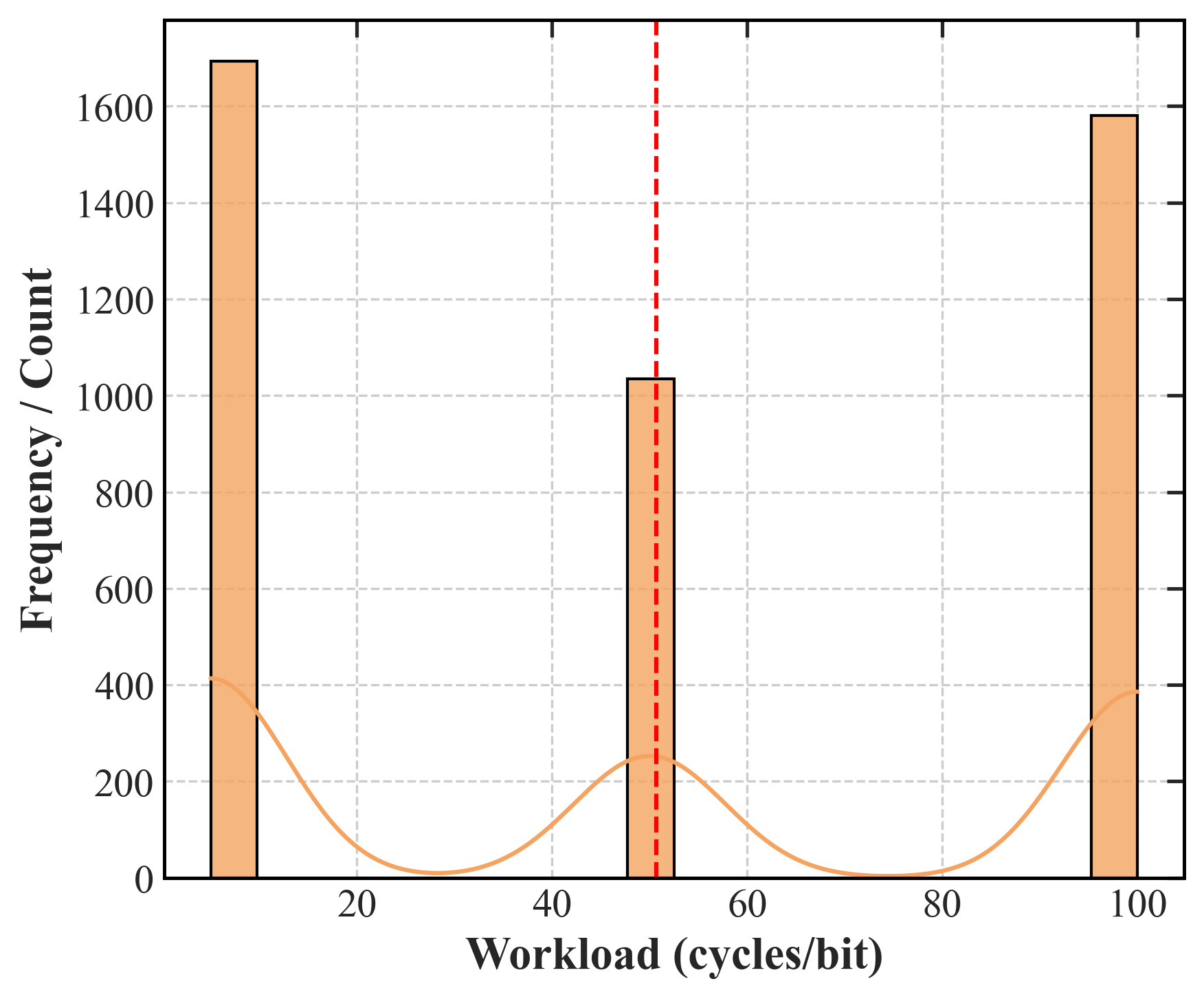}%
    }
    \caption{Workload characteristics.}
    \label{fig:task_dist}
\end{figure}

\add{As depicted in \figurename~\ref{fig:task_dist}, this configuration 
produces a bimodal data-size distribution (concentrated around small-payload 
mission-critical tasks and large-payload raw sensor data) and a tri-modal 
computational-density distribution (corresponding to the three operational 
profiles). Defining the coefficient of variation as $\mathrm{CV} = \sigma/\mu$ 
(the ratio of the standard deviation to the mean), we obtain 
$\mathrm{CV}_{\text{size}} = 0.39$ for data sizes and 
$\mathrm{CV}_{\text{work}} = 0.82$ for computational workloads, indicating 
moderate heterogeneity in data volumes and substantial heterogeneity in 
computational workloads. This combination stresses both bandwidth-bound and 
compute-bound routing decisions and exposes the scheduler to a wide range 
of operational regimes.}

\subsubsection{Fault Injection Scenarios}
\begin{figure}[tb!]
    \centering
    \includegraphics[width=\linewidth]{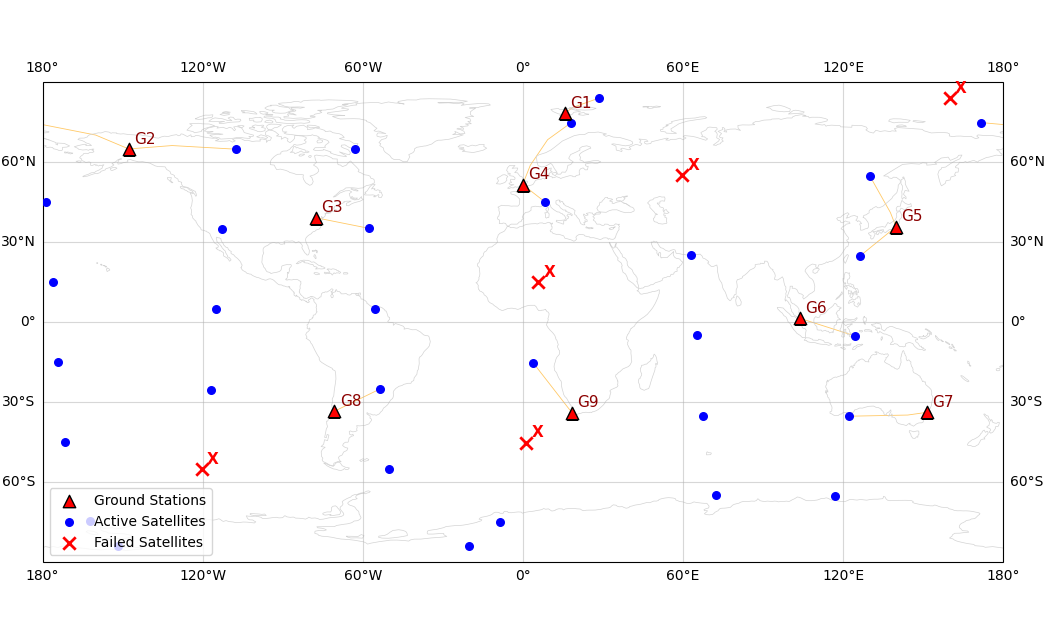}
    \caption{{Global topology snapshot under massive area denial ($t=3600$\:s).}}
    \label{fig:topo_map}
\end{figure}

To evaluate system resilience beyond normal operations, we design a two-stage 
fault injection schedule:
\begin{itemize}[leftmargin=0.3cm, noitemsep]
    \item \textbf{Stage I: Point Failure ($t=3000$ s)}: Satellite \texttt{S5} incurs a permanent node failure, representing a single-point outage that necessitates localized routing reconfiguration.
    \item \textbf{Stage II: Massive Area Denial} ($t=3600$\:s): Five satellites 
    (\texttt{S4, S10, S16, S18, S32}) fail simultaneously. As shown in 
    \figurename~\ref{fig:topo_map}, these failures are distributed across multiple orbital planes, removing approximately $14$\% of active network nodes.
\end{itemize}

\subsubsection{Baselines and Ablation Variants}
To isolate the performance gain of our proposed RAG-driven context-aware reasoning framework \textbf{\sysname}, we compare against two categories of methods. The first comprises routing and resources offloading baselines that differ fundamentally in architecture from \sysname, while the second consists of ablation variants that systematically remove individual design components to validate their respective contributions. The baselines are summarized as below:
\begin{itemize}[leftmargin=0.3cm, noitemsep]
    \item \textbf{QoS-Aware KSP~\cite{SOTA_KSP}:} A purely reactive $K$-Shortest Path routing algorithm incorporating instantaneous queueing delays into edge weights.
    \item \textbf{Static GA~\cite{qi2025ls2moss}:} The baseline GA operating with fixed, static multi-objective weights ($\omega_{lat}=1, \omega_{cong}=1, \omega_{eng}=1$).
       \item \textbf{RAG+KNN:} A non-LLM baseline combining the RAG paradigm~\cite{huang2026ragsurvey} with
continuous regression. At inference time, it computes the cosine similarity between the current network abstract state
and all historical expert weight configurations stored in the
EKB, retrieves the top-$K$ most similar records, and produces
the current preference weight vector $\boldsymbol{\omega}_t =
(\omega_{\text{lat}}, \omega_{\text{cong}},
\omega_{\text{eng}})$ via cosine-similarity-weighted
interpolation over the retrieved three-dimensional weight
vectors.

\item \textbf{RAG+MLP:} A non-LLM baseline using a
supervised Multi-Layer Perceptron trained offline as a
regression surrogate. At each inference step, it first
performs a real-time RAG
retrieval~\cite{huang2026ragsurvey} to obtain
the Top-$1$ most similar expert weight configuration from the
EKB, then feeds a concatenated augmented vector composed of the current network state features and the retrieved weight features into the MLP to directly predict $\boldsymbol{\omega}_t = (\omega_{\text{lat}},
\omega_{\text{cong}}, \omega_{\text{eng}})$, serving as an
ultra-fast surrogate that replaces the LLM in the preference
weight generation stage.
\add{\item \textbf{O2O-PPO}~\cite{SOTA_PPO}: An offline-to-online DRL 
baseline. A policy network is first initialized via Behavior Cloning on 
expert (state, preference) pairs $(\mathbf{s}_{\text{off}}, 
\boldsymbol{\omega}_{\text{off}})$ drawn from the EKB, then refined online 
via Proximal Policy Optimization (PPO) to map $\mathbf{s}_t$ directly to 
$\boldsymbol{\omega}_t$. This baseline replaces both the RAG retrieval and 
LLM inference stages of \sysname{} with an end-to-end learned policy, 
isolating the contribution of retrieval-augmented LLM-based inference 
against a fully learning-based alternative trained on the same expert data.}
\end{itemize}

Two ablation variants are summarized as below:
\begin{itemize}[leftmargin=0.3cm, noitemsep]
    \item \textbf{Naive LLM (Zero-Shot)~\cite{SOTA_ZEROSHOT}:} An LLM agent directly reasoning over raw telemetry prompts without historical priors or chain-of-thought, outputting weights zero-shot.
    \item \textbf{NoRAG (CoT-Reasoning):} It is an LLM-based method developed from~\cite{SOTA_CoT} enhanced with physics-informed CoT prompting (explicitly diagnosing compute vs. bandwidth regimes) but strictly operating in a closed-book manner without external retrieval.

\end{itemize}

To ensure absolute evaluation fairness, all parameterized learning and generative baselines share the exact same lower-level fidelity-aware genetic scheduler. These upper-level agents are exclusively tasked with deducing the preference weight vector $\boldsymbol{\omega}_t$, leaving the physical collision-free scheduling to the identical GA executor. Thus, any macro-performance gain is strictly attributed to superior hyperparameter orchestration, not the underlying combinatorial solver.

\subsubsection{Implementation Settings}~\label{subsubsec:implementation}
The simulation environment is implemented in Python 3.9 using \texttt{NetworkX}{\footnote{\url{https://networkx.org/en/}}} for graph-theoretic operations and \texttt{Skyfield} for high-fidelity orbital mechanics. The cognitive agent interfaces asynchronously with a DeepSeek-R1-Distill (1.5B) model. In the simulation, the cognitive agent interfaces asynchronously 
via a cloud-based API using DeepSeek-R1-Distill (1.5B) to manage 
concurrent reasoning tasks. For hardware viability projection, we 
consider a lightweight 0.7B onboard vision-language model deployed 
on the NX1 accelerator, with a decoding speed of 
$\nu_{\mathrm{dec}} \approx 23.45$\:tokens/s as measured on the 
target hardware.

Specifically, to validate the real-world viability of the proposed \sysname, we perform a \textit{hardware-anchored latency projection} based on the NX1 onboard computer (Table~\ref{tab:hardware_specs}), which delivers $248$\:TOPS of INT8 throughput. Based on established edge-accelerator benchmarks for 1.5B models, we adopt a realistic inference model with a generation speed of $\nu_{\mathrm{dec}} \approx 23.45$\:tokens/s.

\begin{table}[tb!]
\centering
\caption{Specifications of the reference onboard computing platform (NX1)}
\label{tab:hardware_specs}
\renewcommand{\arraystretch}{1.1}
\begin{tabular}{l|l}
\hline
\textbf{Specification} & \textbf{Parameter Value} \\ \hline
Architecture & Heterogeneous (GPU-SoC + FPGA-SoC) \\
AI computing power & 248 TOPS (INT8) \\
GPU memory & 64 GB (with ECC protection) \\
Power consumption & 93.5 W (Typical) / 153 W (Peak) \\
\hline
\end{tabular}
\end{table}

Deploying LLMs for direct network control introduces risks of semantic hallucination, where generated weights might lead to physical infeasibility. To ensure absolute operational safety, the lower-level execution plane implements two deterministic guardrails:
\begin{itemize}[leftmargin=0.3cm, noitemsep]
    \item \textbf{Weight Clipping}: LLM-generated preference weights are mathematically clipped to a strict operational bounding box $\boldsymbol{\omega} \in [0.5, 100]$. This prevents anomalous zero or negative penalties that would invalidate the optimization objective.
    \item \textbf{Congestion Failsafe (Physics Override)}: If the instantaneous CPU queue backlog exceeds the hardware buffer tolerance, the system triggers a deterministic override. This forcefully escalates the energy penalty $\omega_{\mathrm{eng}}$ to its maximum, mandating direct multi-hop transmission and bypassing further onboard processing to prevent hardware saturation.
\end{itemize}

These heuristic boundaries ensure the framework remains strictly bounded by physical hardware limits, even if the upper-level cognitive agent extrapolates unseen or extreme scenarios.

\subsection{Overall Performance Analysis}
\label{subsec:main_results}
\begin{figure*}[!t]
    \centering
    \subfloat[Packet loss rate\label{fig:loss_rate:a}]{%
        \includegraphics[width=0.32\linewidth]{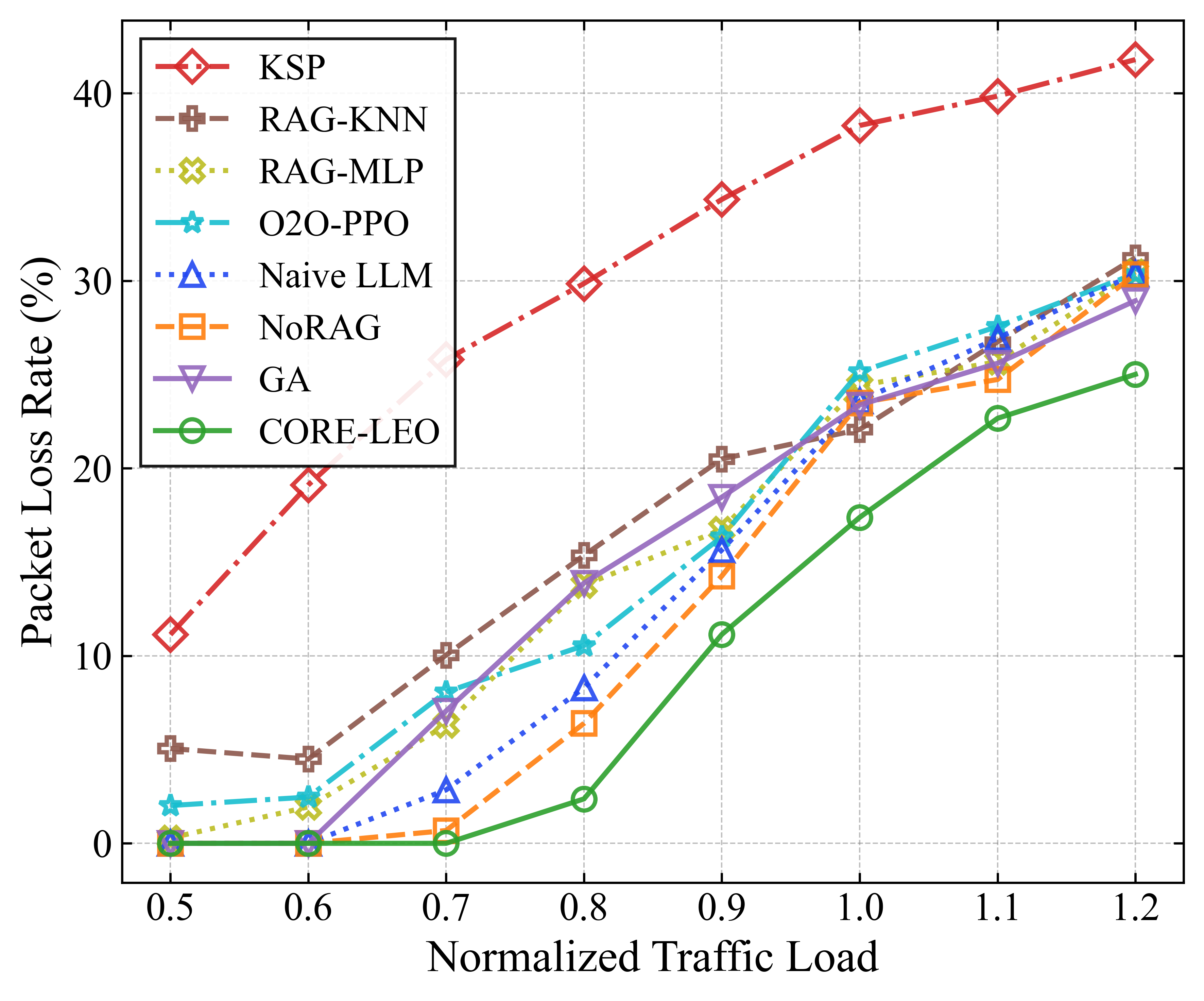}%
    }
    \hfil
    \subfloat[System throughput\label{fig:throughput:b}]{%

        \includegraphics[width=0.32\linewidth]{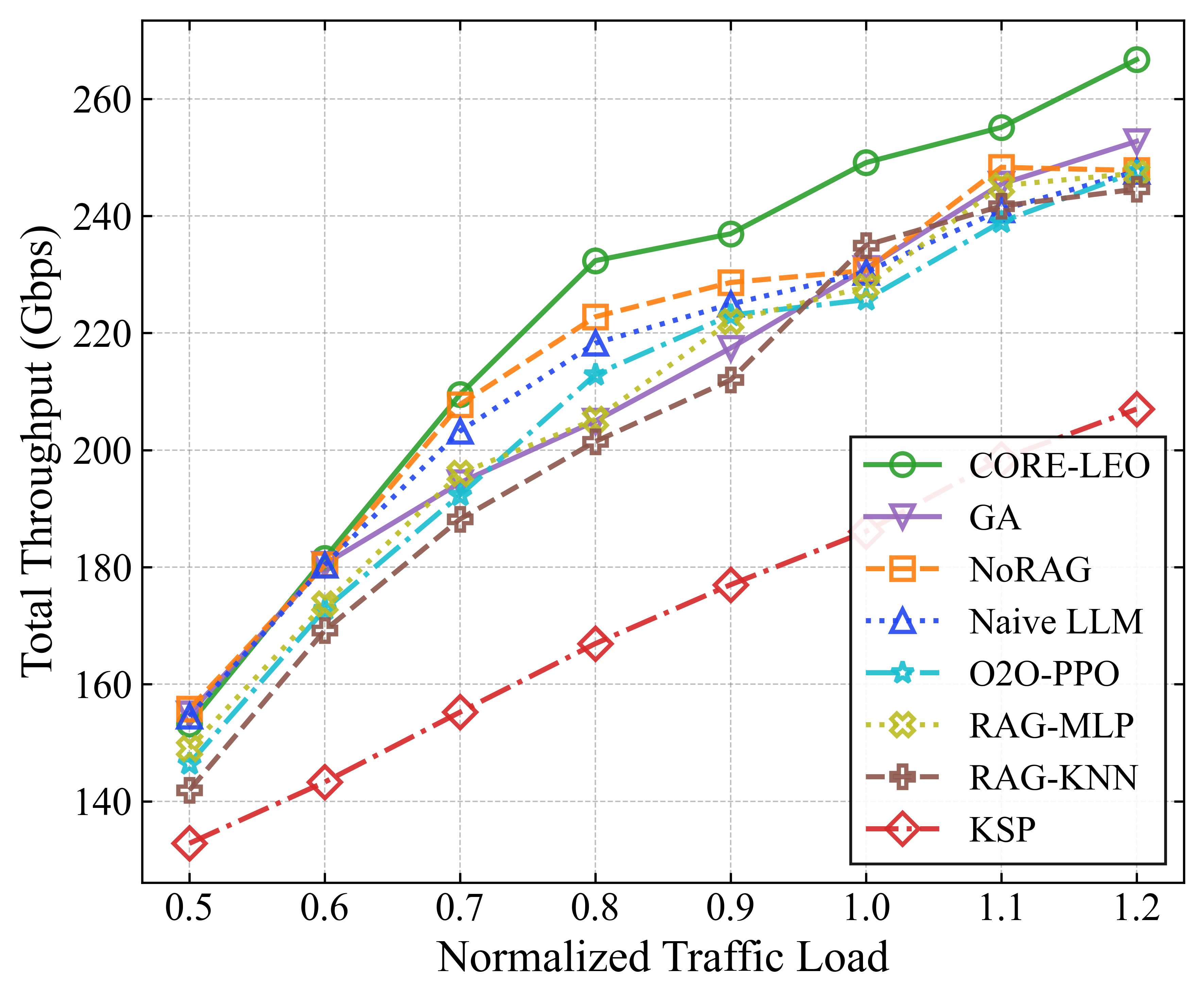}%
    }
    \hfil
    \subfloat[Average latency\label{fig:latency:c}]{%
        \includegraphics[width=0.32\linewidth]{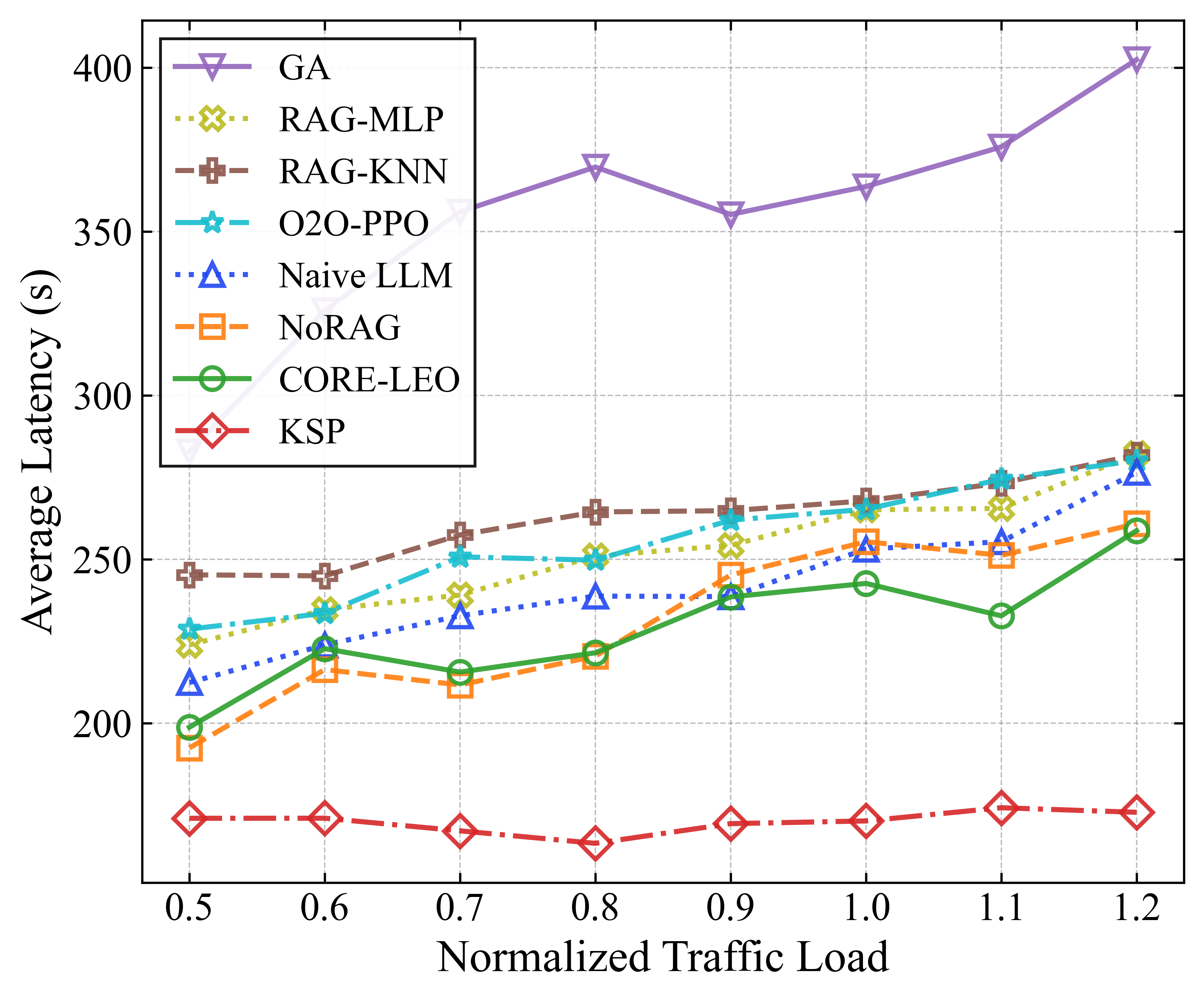}%
    }
    \caption{Overall performance under different traffic loads.}
    \label{fig:load_analysis}
\end{figure*}

We evaluate the overall performance metrics over the full simulation duration ($T=5400$\:s). Table~\ref{tab:performance_snapshot} provides a numerical snapshot of the system under the standard saturated workload ($1\times$), while \figurename~\ref{fig:load_analysis} illustrates the scalability and temporal dynamics of various methods under varying traffic intensities.

\begin{table*}[t]
\centering
\caption{Performance snapshot at standard saturated load (1$\times$).}
\label{tab:performance_snapshot}
\renewcommand{\arraystretch}{1.2}
\begin{threeparttable}
\begin{tabular}{lcccccc}
\toprule
\textbf{Method} & \textbf{Loss Rate (\%)} $\downarrow$ & \textbf{Throughput (Gbps)} $\uparrow$ & \textbf{Avg Latency (s)} $\downarrow$ & \textbf{Energy Eff. (Mb/J)} $\uparrow$ & \textbf{Avg Hops} & \textbf{JFI} $\uparrow$ \\
\midrule
\multicolumn{7}{l}{\textit{Baselines}} \\
\midrule
KSP        & 38.3 & 186.1 & 170.2\tnote{*} & 17.2\tnote{*} & 5.18 & 0.466 \\
Static GA  & 23.4 & 231.0 & 363.8          & 11.5          & 5.17 & 0.349 \\
RAG + KNN  & 22.1 & 235.0 & 267.8          &  9.7          & 4.97 & 0.526 \\
RAG + MLP  & 24.4 & 227.9 & 265.1          &  9.6          & 4.91 & 0.494 \\
O2O-PPO    & 25.1 & 225.7 & 265.3          &  9.0          & 4.81 & 0.499 \\
\midrule
\multicolumn{7}{l}{\textit{Ablation Variants}} \\
\midrule
Naive LLM  & 23.6 & 230.4 & 253.1 & 10.1 & 5.18 & 0.517 \\
NoRAG      & 23.5 & 230.7 & 255.4 & 10.1 & 5.18 & 0.496 \\
\midrule
\textbf{\sysname{}} & \textbf{17.4} & \textbf{249.1} & \textbf{242.7} & \textbf{11.7} & 5.29 & \textbf{0.537} \\
\bottomrule
\end{tabular}
\begin{tablenotes}[flushleft]
\footnotesize
\item Arrows in column headers indicate optimization direction: $\downarrow$ lower is better, $\uparrow$ higher is better. Bold values mark the best result in each column among methods with $<25\%$ loss rate. \textbf{Avg Hops} reports the average path length and has no preferred direction.
\item[*] KSP metrics are statistically skewed by severe ingress-level traffic shedding (38.3\% loss): only short-path flows survive, biasing both latency and energy-efficiency downward. These two values are excluded from the bold comparison.
\end{tablenotes}
\end{threeparttable}
\end{table*}

\subsubsection{Performance Snapshot under Standard Saturated Load}
As shown in Table~\ref{tab:performance_snapshot}, under the standard saturated load ($1\times$), the proposed \sysname achieves a packet loss rate of $17.4$\%, outperforming all baselines. This represents a relative reduction of $30.7$\% over the O2O-PPO baseline (from $25.1$\% to $17.4$\%), and improves upon both learning-based retrieval baselines (RAG+KNN at $22.1$\% and RAG+MLP at $24.4$\%), yielding an aggregate throughput of $249.1$\:Gbps.

We analyze these results across two operational dimensions:
\begin{itemize}[leftmargin=0.3cm, noitemsep]
    \item \textbf{Throughput and Load Balancing:} Beyond packet loss, the proposed RAG-enhanced framework achieves the highest Jain's Fairness Index ($JFI = 0.537$)~\cite{jain1984fairness}, compared to $0.349$ for Static GA. This indicates that the LLM-generated routing weights distribute traffic more uniformly across the network, effectively mitigating localized congestion without monopolizing critical inter-plane links.

    \item \textbf{Routing Behavior and Influence of Route Selection Policies:} \sysname exhibits the highest average path length at $5.29$ hops. Rather than strictly adhering to topologically shortest paths, the agent intelligently trades marginal increases in propagation distance for congestion avoidance, recovering traffic that shortest-path routing would otherwise drop, and achieving a stable end-to-end latency of $242.7$\:s. In contrast, while KSP records a numerically lower average latency ($170.2$\:s) and higher energy efficiency ($17.2$\:Mb/J), this is primarily a consequence of its $38.3$\% packet drop rate, which disproportionately filters out multi-hop flows. Similarly, the learning-based retrieval baselines (RAG+KNN and RAG+MLP) exhibit the lowest average hop counts ($4.97$ and $4.91$). Without the capacity for context-aware reasoning, these models tend to select topologically shorter paths, accelerating local bottleneck saturation under high traffic loads.
\end{itemize}
    
\subsubsection{Scalability Across Varying Traffic Regimes}
To evaluate system robustness, we vary the normalized traffic load from $0.5$ to $1.2$, as depicted in \figurename~\ref{fig:load_analysis}. Under light load conditions ($0.5 - 0.6$), network resources remain abundant, and all AI-based algorithms exhibit comparable performance. However, as the network transitions into the \textbf{critical saturation regime} ($0.7 - 1.0$), the proposed \sysname significantly widens the performance gap. By effectively exploiting fragmented bandwidth and computational resources, it sustains the highest throughput growth (\figurename~\ref{fig:load_analysis}b) and the most gradual increase in packet loss rate (\figurename~\ref{fig:load_analysis}a).At extreme overload states (1.2), the loss rates of all agents begin to converge as the absolute physical capacity boundaries of the LEO satellite network are reached, \add{consistent with the saturation regime predicted by classical queueing analysis: when the aggregate offered load approaches the network's total service capacity, even optimal scheduling cannot prevent queue buildup, and inter-method differences diminish}.

\subsubsection{Latency-Reliability Trade-off and Pareto Analysis}
\begin{figure}[t]
    \centering
    \includegraphics[width=\columnwidth]{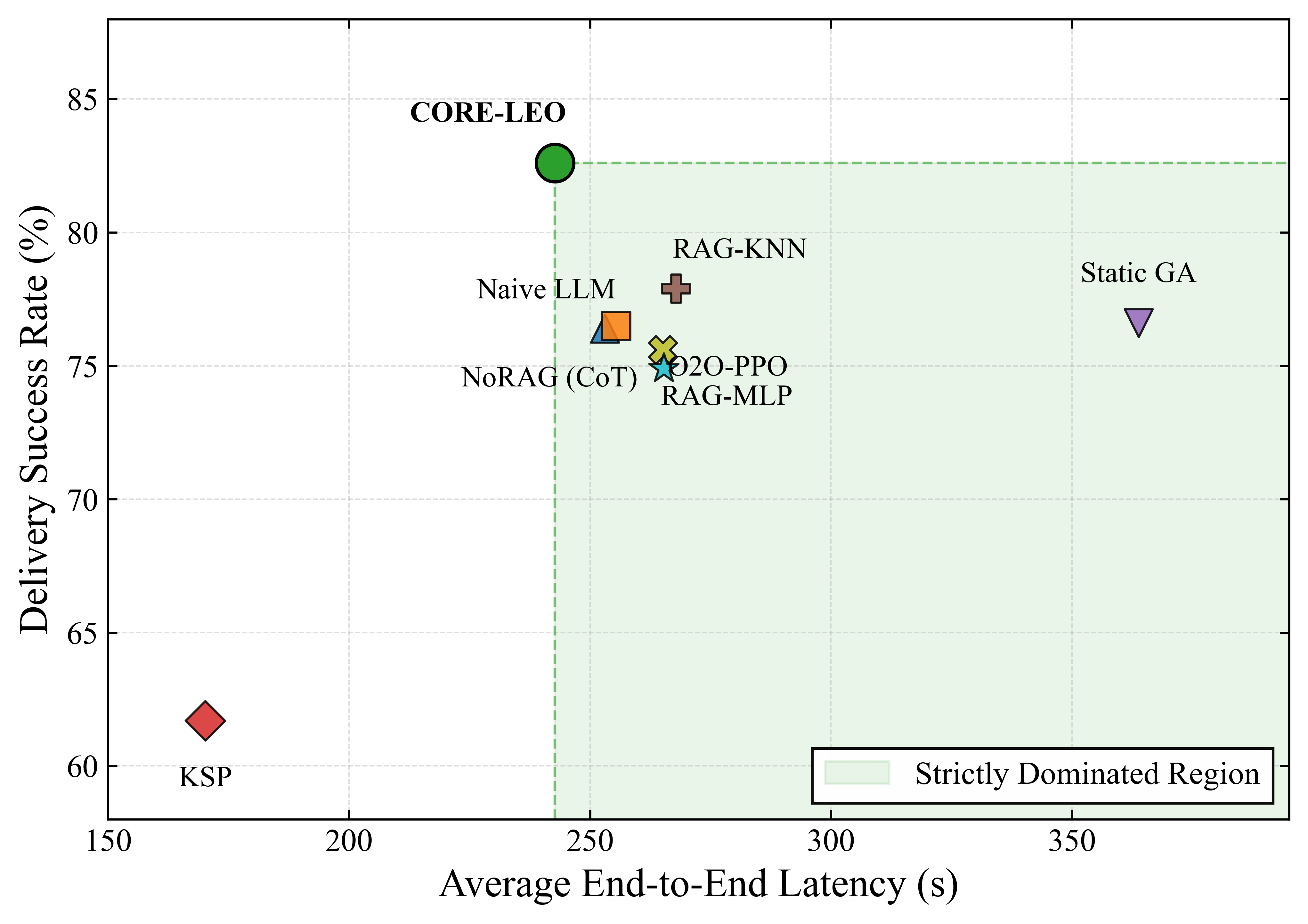}
    \caption{Latency-reliability Pareto evaluation under standard saturated load.}
    \label{fig:pareto_tradeoff}

\end{figure}

The complex relationship between latency and service reliability is plotted in \figurename~\ref{fig:pareto_tradeoff}, which maps the delivery success rate \add{(in percent, defined as $100\% - \mathrm{PLR}$, where PLR is the packet loss rate)} against average end-to-end latency. As mentioned above, the KSP baseline's \add{seemingly low latency ($170.2$\:s, marked with an asterisk in Table~\ref{tab:performance_snapshot})} is a consequence of dropping over $38$\% of packets\add{: only the easy, short-path traffic survives selection, and the long-haul flows that would have inflated the average are systematically discarded at ingress}. Conversely, methods enforcing strict delivery without adaptive routing incur extreme queueing delays. Static GA yields an average latency exceeding $363$\:s due to packet accumulation in congested CPU buffers.
The green shaded area in \figurename~\ref{fig:pareto_tradeoff} identifies the strictly dominated region. The generative, heuristic, and RL-based baselines (Naive LLM, NoRAG, O2O-PPO, and learning-based retrieval baselines) cluster within this zone, indicating that their routing configurations incur higher delays without proportional gains in delivery success. In contrast, \sysname establishes a distinct operational point on the Pareto frontier. By deliberately introducing a controlled increase in propagation delay, directing traffic across longer, spatially disjoint paths to bypass local bottlenecks, the framework operates safely outside the dominated region. This strategic detour mechanism achieves a superior delivery success rate of approximately $82.6$\%, preserving multi-hop traffic that traditional algorithms typically discard under stringent topological constraints.

\subsubsection{Energy Efficiency and Computation-Communication Balance}
Returning to the data in Table~\ref{tab:performance_snapshot}, \sysname achieves a network energy efficiency of $11.7$\:Mb/J, the highest among all AI-augmented baselines. Crucially, the orchestrator successfully avoids inefficient operating regimes where the queueing latency at overloaded satellite CPUs outweighs the transmission latency saved by data compression. By dynamically leveraging the \textit{deterministic physics override} mechanism (detailed in Section~\ref{subsec:ul}), the agent identifies compute-bound orbital regimes and autonomously transitions to direct transmission, conserving satellite battery reserves while preventing catastrophic queue overflow.


\subsection{Resilience Analysis: Dynamic Response to Failures}
\label{subsec:resilience}
We evaluate the temporal stability and autonomous recovery capabilities of the orchestration framework under dynamic fault injections. \figurename~\ref{fig:cumulative_delivery} tracks the cumulative volume of successfully delivered data over the simulation period.

\subsubsection{Dynamic Response to Massive Area Denial}
\begin{figure}[t]
\centering
 \includegraphics[width=\columnwidth]{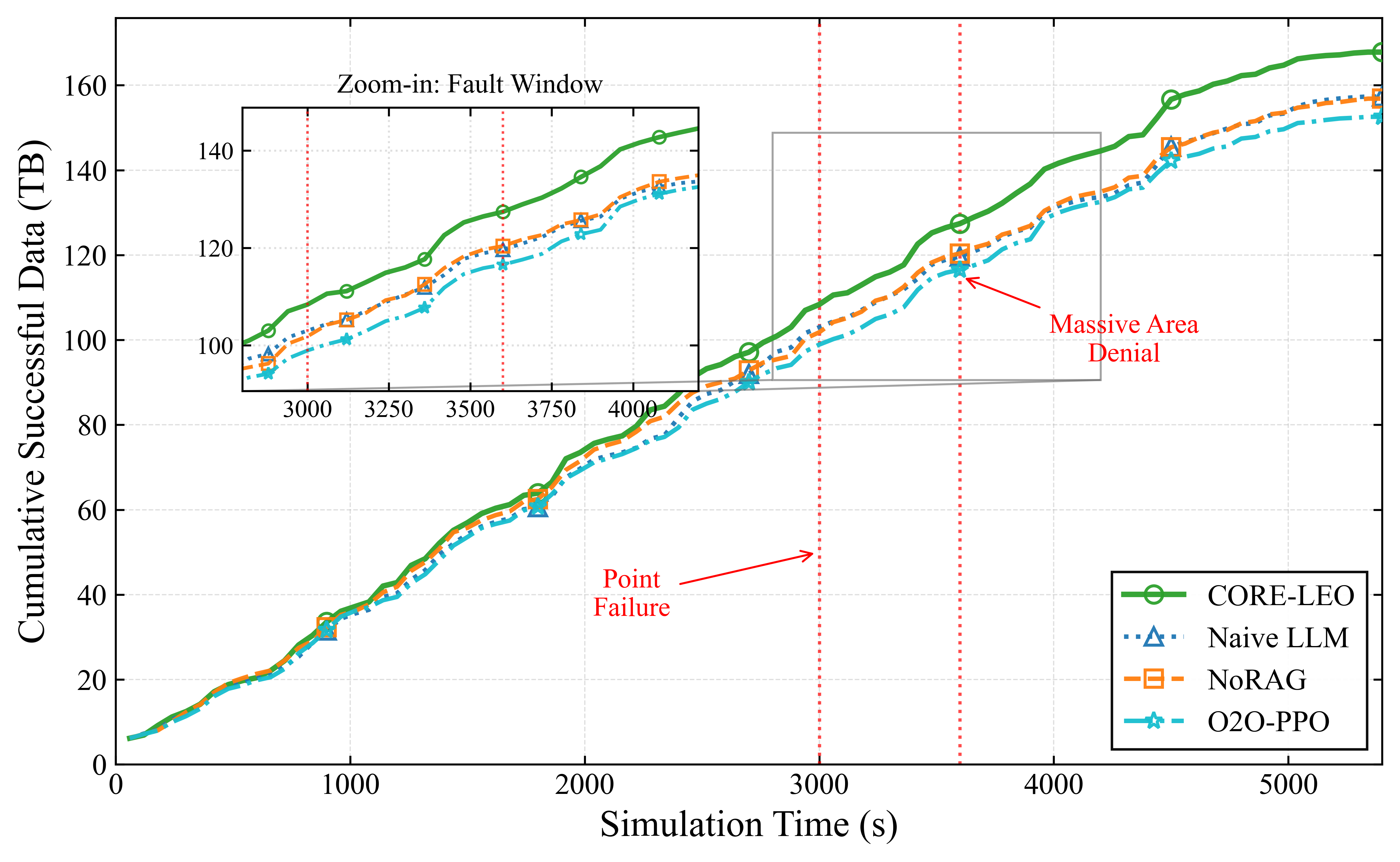}
\caption{{Cumulative data delivery under dynamic topology disruptions.}}
\label{fig:cumulative_delivery}
\end{figure}

The zoom-in plot in \figurename~\ref{fig:cumulative_delivery} focuses on the critical disruption window ($t \in [3000, 4000]$\:s), capturing the network's transient response to multiple concurrent topological failures.
\begin{itemize}[leftmargin=0.3cm, noitemsep]
    \item \textbf{Baseline Degradation}: Immediately following a 5-node failure event at $t=3600$\:s (which compounds an earlier point failure for a total of $6$ disabled satellites), the Naive (\textit{yellow dashed curve}), NoRAG (\textit{orange curve}), and O2O-PPO (\textit{blue dash-dot curve}) baselines all exhibit a marked reduction in delivery slope. This reduction in instantaneous throughput indicates the inability of retrieval-free baselines to rapidly recompute viable multi-hop routes around abrupt topological disruptions. The absence of historical structural priors yields localized queue delays, elevated packet drops, and a prolonged period of reduced delivery rates.

    \item \textbf{RAG-Driven Resilience}: Conversely, \sysname (\textit{green solid curve}) maintains a near-constant data delivery rate throughout the cascading disruption. By mapping the instantaneous network telemetry to similar historical topology states stored within the offline EKB ($\mathcal{H}_{\mathrm{RAG}}$), the cognitive agent retrieves pre-validated fallback strategies. This retrieval mechanism enables our framework to proactively redistribute traffic across surviving orbital planes, ensuring service continuity even when approximately $16.7$\% of the network's routing capacity is instantaneously disabled.
\end{itemize}

\subsubsection{Throughput Composition and Network Utilization Efficiency}
\label{subsubsec:throughput_composition}
\figurename~\ref{fig:priority} disaggregates cumulative throughput by priority tier, isolating mission-critical traffic ($p \in \{3, 5\}$) from best-effort telemetry ($p=1$). This breakdown reveals how different routing agents arbitrate between competing traffic classes under heterogeneous load, exposing four distinct behavioral regimes. The internal link-level mechanisms underlying these outcomes are further examined in Section~\ref{subsubsec:queue_dynamics}.
\begin{itemize}[leftmargin=0.3cm, noitemsep]
\item \textbf{KSP (Traffic Shedding):}
KSP's leading mission-critical throughput ($120.3$\:Gbps) is not a product of efficient load management, but an artifact of strict admission control. By discarding the majority of best-effort flows at the network ingress, it delivers only $65.8$\:Gbps of priority-1 traffic and achieves the lowest JFI ($0.466$) among all methods, confirming that its apparent high-priority advantage \add{is achieved at the cost of severe inter-class unfairness}.

\item \textbf{Numerical Baselines (Path Inflexibility):}
RAG+KNN and RAG+MLP leverage the historical experience base but are confined to statistical function approximation, lacking context-aware reasoning over dynamic topology states. Under localized congestion, this limitation produces conservative, averaged routing weights that are insufficient to trigger decisive spatial detours. Both baselines consequently exhibit \textit{path inflexibility}, a persistent tendency to concentrate mixed-priority traffic along shortest-path corridors regardless of instantaneous link load, leading to inevitable buffer saturation and cascading packet loss at primary bottleneck nodes.

\item \textbf{Generative and DRL Baselines (Routing Weight Instability):}
Naive LLM, NoRAG, and O2O-PPO possess context-aware reasoning or adaptive policy capabilities but operate without retrieval-augmented physical anchors. In a dynamic $36$-node network, inference without retrieval grounding struggles to accurately model graph-structured topology constraints, frequently producing routing hyperparameters that fail to converge to physical optima. This \textit{routing weight instability} manifests as degraded throughput across both priority tiers, i.e., Naive LLM ($230.4$\:Gbps), NoRAG ($230.7$\:Gbps), and O2O-PPO ($225.7$\:Gbps), with JFI values of $0.517$, $0.496$, and $0.499$, respectively, indicating residual inter-class imbalance despite the absence of explicit admission control.

\item \textbf{\sysname (Spatial Redistribution):}
\add{By conditioning the LLM 
inference on retrieved historical priors that are themselves derived from 
physically feasible Pareto-optimal configurations}, \sysname{} achieves the 
highest cumulative throughput (249.1 Gbps) while simultaneously attaining 
the best inter-class fairness ($\mathrm{JFI} = 0.537$).
\end{itemize}
\begin{figure}[t]
\centering
\includegraphics[width=\columnwidth]{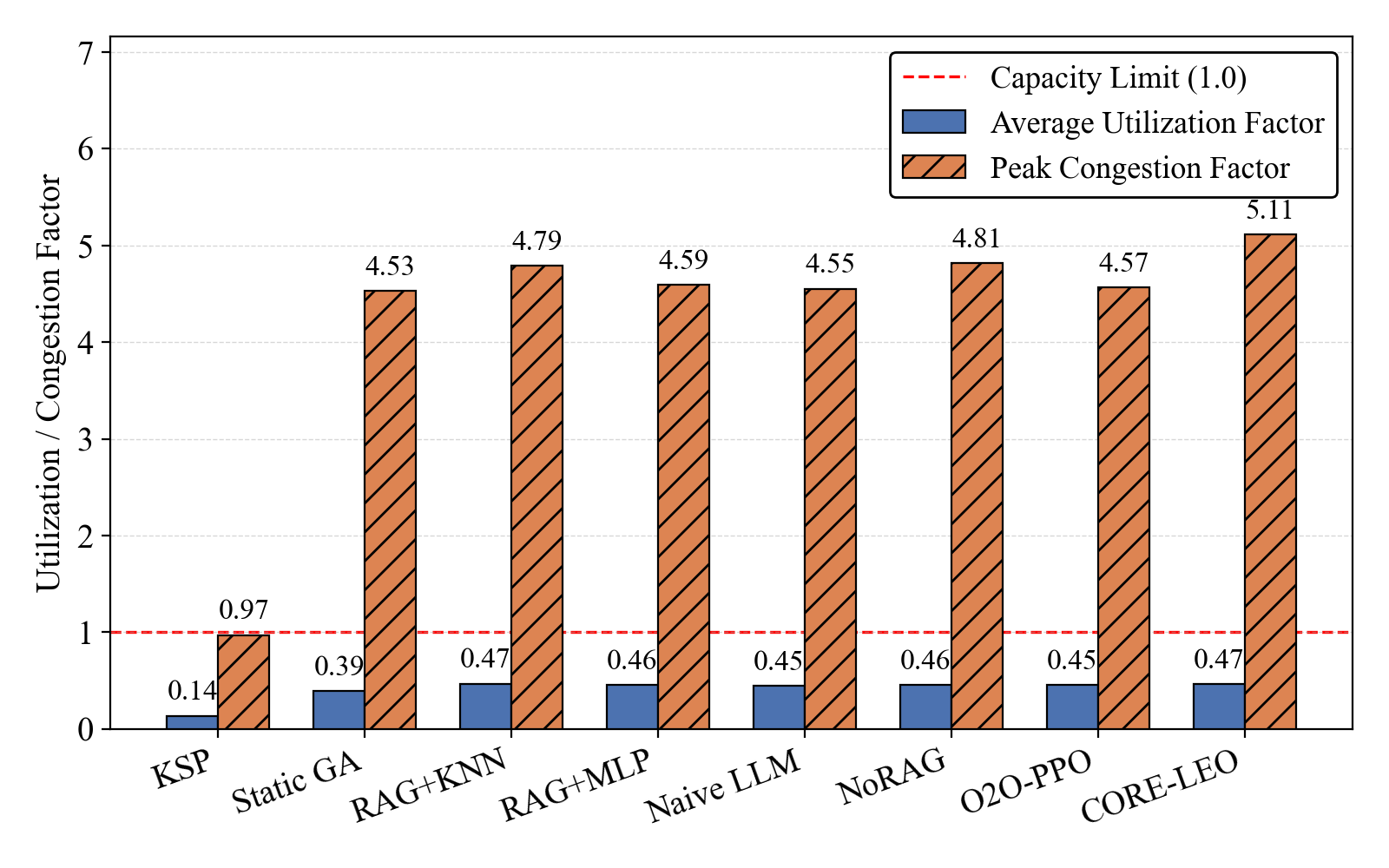}
\caption{{Average utilization vs. peak link congestion factor under saturated load.}}
\label{fig:link_utilization}
\end{figure}
\subsubsection{\add{Link Resource Allocation and Congestion Patterns}}
\label{subsubsec:queue_dynamics}
\begin{figure}[t]
    \centering
    \includegraphics[width=\columnwidth]{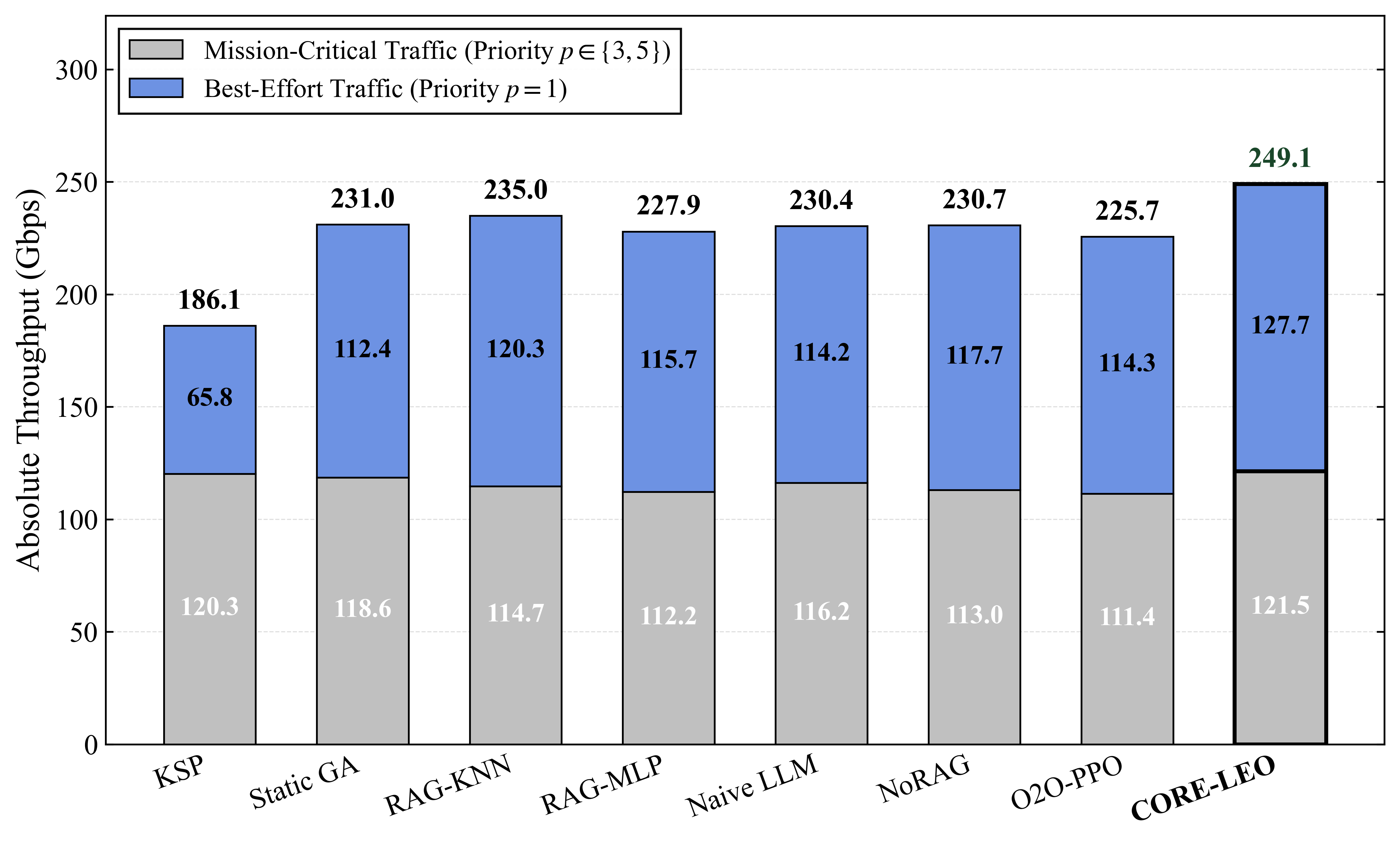}
    \caption{{Absolute throughput composition across priority tiers.}}
    \label{fig:priority}
\end{figure}

\figurename~\ref{fig:link_utilization} examines the internal resource allocation profile of each routing strategy through two complementary lenses. The \textit{average utilization factor} captures the spatial coverage of active ISL paths across the network, while the \textit{peak congestion factor} indicates whether traffic successfully propagates to terminal sink links (SGLs). Interpreted jointly with the packet loss rates reported in Table~\ref{tab:performance_snapshot}, these metrics distinguish \textit{sink-driven congestion}, in which queue accumulation at physical sinks reflects high delivery volume, from \textit{upstream truncation} caused by premature queue saturation along primary corridors.
\begin{itemize}[leftmargin=0.3cm, noitemsep]
\item \textbf{KSP (Utilization Suppression):}
KSP's average utilization factor ($0.14$) is the lowest across all methods, reflecting aggressive ingress admission control and inefficient resource management. By shedding best-effort traffic before network entry, it artificially suppresses ISL queue depth, yielding a peak link congestion factor of $0.97$, the only value below the capacity limit. Its $38.3$\% loss rate is thus attributable to aggressive ingress-level traffic shedding.

\item \textbf{Generative and DRL Baselines (Upstream Truncation):}
Naive LLM, NoRAG, and O2O-PPO all achieve moderate link congestion factors ($4.55$, $4.81$, and $4.57$, respectively), which may appear to suggest effective sink utilization. Their packet loss rates ($23.6$\%, $23.5$\%, and $25.1$\%) reveal a different mechanism. Without spatial detour capabilities, these agents concentrate traffic on shortest-path corridors, saturating upstream queues before flows reach the terminal SGLs. The sink-level congestion observed is not a sign of high delivery volume, but instead represents \textit{upstream truncation} that exhausts buffer resources without improving throughput.

\item \textbf{Numerical Baselines (Spatial Constriction):}
RAG+KNN and RAG+MLP exhibit a distinct pattern: despite average hop counts of $4.97$ and $4.91$, lower than most other methods, their average utilization factors ($0.47$ and $0.46$) are comparable to \sysname{}. This apparent inconsistency indicates that their ISL engagement is spatially constricted to a narrow set of high-load corridors instead of being distributed across the network. The resulting bottleneck concentration manifests as elevated loss rates, particularly for RAG+MLP ($24.4$\%), undermining the throughput gains achievable through broader spatial distribution.

\item \textbf{\sysname{} (Sink-Driven Congestion):}
\sysname{} achieves the highest link congestion factor ($5.11$) alongside the lowest packet loss rate ($17.4$\%), a combination that jointly characterizes \textit{sink-driven congestion}. The elevated average hop count ($5.29$), the highest among all methods, confirms that the framework actively engages peripheral ISL paths to spatially redistribute traffic load, utilizing the network's distributed buffering capacity instead of concentrating demand at primary corridors. Consequently, queue accumulation at the terminal SGLs reflects genuine high-volume delivery, directly accounting for the $249.1$\:Gbps aggregate throughput reported in \figurename~\ref{fig:priority}.
\end{itemize}

\subsection{Feasibility Analysis}
\label{subsec:feasibility}
Finally, we validate the computational cost of the proposed bi-level architecture on the reference onboard platform (Table~\ref{tab:hardware_specs}).
The total per-decision-cycle latency comprises two stages. First, the RAG-driven LLM inference incurs a median latency of approximately $18$\:s, derived from the measured decoding throughput of the onboard 0.7B model ($\nu_{\mathrm{dec}} \approx 23.45$\:tokens/s) on the NX1 accelerator. Second, the lower-level genetic scheduler, which subsumes candidate path generation and GA optimization within a single execution pass, requires an average of $16$\:s per round across the full $5400$\:s simulation under standard saturated load ($1\times$). The aggregate end-to-end decision latency is therefore approximately $18 + 16 = 34$\:s, consuming roughly $56.7$\% of the $T_D = 60$\:s topological contact window defined in Section~\ref{subsubsec:implementation}, and leaving a conservative margin of $26$\:s. Although inherently higher than static mathematical heuristics, this overhead remains well within the operational boundary, providing sufficient headroom to absorb transient load spikes or extended inference under extreme congestion scenarios. Given that the RAG-driven cognitive policy yields a packet loss reduction 
of $30.7$\% relative to O2O-PPO, the computational overhead is justified in practice.

\section{Conclusion}~\label{sec:conclusion}
In this paper, we propose \sysname, a bi-level cognitive orchestration 
framework designed to mitigate spatio-temporal resource fragmentation in LEO satellite networks. By decoupling macroscopic network intent from microscopic physical execution, the architecture leverages a RAG-enhanced LLM as a cognitive leader. This upper-level agent dynamically infers preference weights based on historical topological priors, which subsequently guide a deterministic, gap-aware genetic scheduler at the lower level to ensure collision-free, physics-compliant task execution. \add{Empirical results on a high-fidelity Walker-Delta testbed show that \sysname reduces packet loss by $30.7$\%, improves energy efficiency by $30$\%, and lowers end-to-end latency by $8.5$\% against a competitive learning-based baseline, with robust performance retained under node-failure scenarios.} \add{Our future work will extend \sysname to a multi-agent framework that coordinates intra- and inter-domain routing across mega-constellations.}

\bibliographystyle{IEEEtran}
\bibliography{refs}

\end{document}